\documentclass[longbibliography,physrev,aps,twocolumn]{revtex4-2}
\usepackage{graphicx,epsfig}
\usepackage{times}
\usepackage{amsmath,amssymb,wasysym}
\usepackage{dsfont} 
\usepackage{color}
\usepackage[linktocpage=true,colorlinks=true, pdfborder={0 0 0},linkcolor=blue,citecolor=blue,urlcolor=blue,anchorcolor=black,linkcolor=blue]{hyperref}

\newcommand{\op}[1]{\hat{#1}}

\newcommand{\normV}[1]{||{#1}||_2}
\newcommand{\normM}[1]{||{#1}||}
\newcommand{\normspec}[1]{||{#1}||_{\text{2}}}

\newcommand{\ident}{\mathds{1}}
\newcommand{\C}{\mathbb{C}}

\newcommand{\R}{\mathbb{R}}

\newcommand{\quoting}[1]{``#1''}
\newcommand{\vc}[1]{\vec{#1}}
\newcommand{\rem}[1]{}

\newcommand{\imagc}[1]{\text{Im}\,#1}

\newcommand{\real}[1]{\text{Re}(#1)}

\newcommand{\oHami}{\op{\Gamma}}

\newcommand{\bra}[1]{\langle#1|}
\newcommand{\ket}[1]{|#1\rangle}

\newcommand{\braket}[2]{\langle#1|#2\rangle}

\newcommand{\prodFro}[2]{\langle#1,#2\rangle_{\text{F}}}
\newcommand{\normFro}[1]{||#1||_{\text{F}}}
\newcommand{\rca}{\xi}

\newcommand{\rcb}{\zeta}

\DeclareMathOperator{\trace}{Tr}

\newcommand{\Hs}{\op{H}_0}
\newcommand{\Hp}{\op{H}_1}
\newcommand{\Ha}{\op{H}}

\newcommand{\pf}{\omega}
\newcommand{\order}{n}

\newcommand{\PT}{${\cal{PT}}$}
\newcommand{\ev}{E}
\newcommand{\evEP}{\ev_{\text{EP}}}
\newcommand{\freqEP}{\omega_{\text{EP}}}
\newcommand{\state}{\psi}

\newcommand{\HEP}{\op{H}_{\text{EP}}}
\newcommand{\HDP}{\op{H}_{\text{DP}}}
\newcommand{\HDPexp}{\op{H}'_{\text{DP}}}

\newcommand{\evDP}{\ev_{\text{DP}}}
\newcommand{\distance}[2]{d({#1},{#2})}
\newcommand{\dist}{\Delta}
\newcommand{\Henrici}{{\cal D}}
\newcommand{\vg}{v_{\text{g}}}
\newcommand{\kc}{k_{\text{c}}}

\begin{document}

\title{Distance between exceptional points and diabolic points and its implication for the response strength of non-Hermitian systems}
\author{Jan Wiersig}
\affiliation{Institut f{\"u}r Physik, Otto-von-Guericke-Universit{\"a}t Magdeburg, Postfach 4120, D-39016 Magdeburg, Germany}
\email{jan.wiersig@ovgu.de}
\date{\today}
\begin{abstract}
Exceptional points are non-Hermitian degeneracies in open quantum and wave systems at which not only eigenenergies but also the corresponding eigenstates coalesce. This is in strong contrast to degeneracies known from conservative systems, so-called diabolic points, at which only eigenenergies degenerate. Here we connect these two kinds of degeneracies by introducing the concept of the distance of a given exceptional point in matrix space to the set of diabolic points. We prove that this distance determines an upper bound for the response strength of a non-Hermitian system with this exceptional point. A small distance therefore implies a weak spectral response to perturbations and a weak intensity response to excitations. This finding has profound consequences for physical realizations of exceptional points that rely on perturbing a diabolic point. Moreover, we exploit this concept to analyze the limitations of the spectral response strength in passive systems. A number of optical systems are investigated to illustrate the theory.
\end{abstract}
\maketitle

\section{Introduction}
\label{sec:intro}
Open quantum and wave dynamics and their exotic degeneracies, so-called exceptional points (EPs), have gained substantial attention in recent years, primarily in optics and photonics~\cite{MA19,ORN19}. At an EP of order~$\order$ exactly~$\order$ eigenenergies (eigenfrequencies) and the corresponding energy eigenstates (modes) coalesce~\cite{Kato66,Heiss00,Berry04,Heiss04,MA19}.  This is very different from a conventional degeneracy, known as a diabolic point (DP)~\cite{BW84}, at which only the eigenenergies coalesce. For an EP to exist, the Hamiltonian $\op{H}$ has to be not only non-Hermitian, $\op{H} \neq \op{H}^\dagger$, but also nonnormal, i.e., $[\op{H},\op{H}^\dagger] \neq 0$.

EPs have been observed in numerous experiments in diverse physical systems~\cite{DGH01,DDG04,DFM07,LYM09,POL14,POL16,RZZ19,CKL10,RBM12,GEB15,SKM16,WHL19}. 
Several applications of EPs have been suggested, such as unidirectional lasing operation~\cite{POL16}, orbital angular momentum microlasers~\cite{MZS16}, sources of circularly polarized light~\cite{RMS17}, topological energy transfer between states~\cite{XMJ16,DMB16}, loss-induced suppression of lasing~\cite{POR14}, mode discrimination in multimode laser cavities~\cite{HMH14}, and sensors with enhanced response~\cite{Wiersig14b,COZ17,HHW17,XLK19,LLS19,WLY20,KCE22}.  

The sensing applications of EPs (a review can be found in Ref.~\cite{Wiersig20c}) rely on the strong spectral response to perturbations. A system with an EP$_\order$ shows an energy (frequency) splitting proportional to the $\order$th root of the perturbation strength~$\varepsilon$~\cite{Kato66}, which for sufficiently small perturbations is larger than the linear scaling near a DP.  In Ref.~\cite{Wiersig22} it was shown that the spectral response to perturbations can be characterized by the so-called spectral response strength $\rca$. A large $\rca$ not only indicates a large spectral response to generic perturbations but also a large intensity response to excitations and a large dynamic response to an initial deviation from the EP eigenstate.

One convenient way to create an EP is to start with a DP and to perturb it appropriately~\cite{KKM03,Berry04,WZS21}. For instance, optical modes of ideal whispering-gallery microcavities naturally come as degenerate pairs of clockwise and counterclockwise traveling waves. Such a DP can be converted into an EP by weak local perturbations~\cite{Wiersig11,POL16,Wiersig18b} or weak boundary deformations~\cite{YKW18,KYW18} without adding extra parasitic radiation losses. 
Another example is the efficient transfer of excitations between energy levels, which is usually done as rapid adiabatic passage through a DP. Weakly perturbing it maps this scheme to one based on encircling an EP, which has been demonstrated experimentally using microwave waveguides~\cite{FSD20}. 
Another example where EPs originating from perturbed DPs have been observed is systems of interacting fermions~\cite{SBL22}.
Finally, DPs naturally appear in periodic systems as Dirac or Weyl points in the band structure. These too can be used to create EPs, which has been done for photonic crystals~\cite{ZHI15,LPL16,ZPY18,CSHW19}, Dirac superconductors~\cite{ZS19}, optically biaxial crystals~\cite{RZZ19}, and for a Hubbard model~\cite{RPY21}. 

\begin{figure}[ht]
\includegraphics[width=0.75\columnwidth]{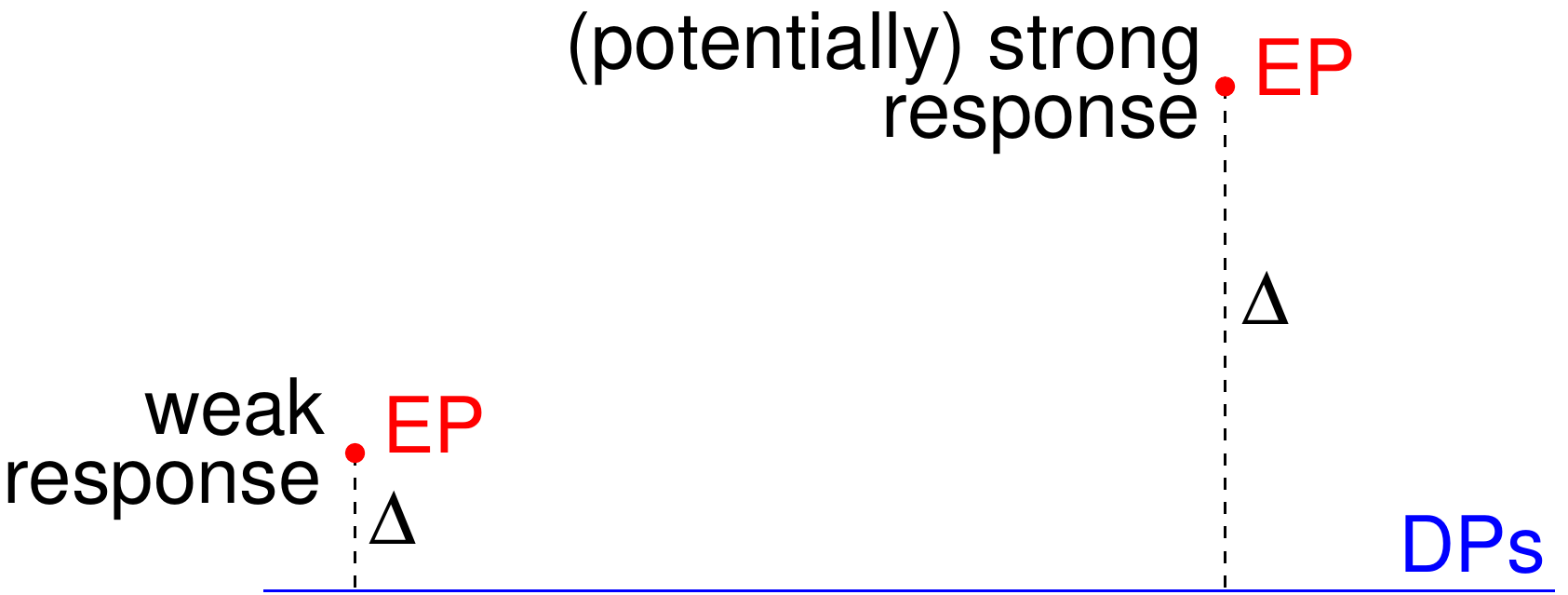}
\caption{Sketch of the distance~$\dist$ (dashed lines) between a given exceptional point (EP) and the set of diabolic points of the same order (DPs, solid line) in matrix space. A key result of this paper is that while a system with an EP which is close to a DP shows only a weak spectral response to perturbations, a system having an EP with a large distance can exhibit a strong spectral response to perturbations.}
\label{fig:sketchdEPDP}
\end{figure}

The aim of this paper is to introduce the concept of the distance~$\dist$ between a given EP$_\order$ and the set of DPs of the same order $\order$. This distance quantifies the difference of the $\order\times\order$ Hamiltonians each exhibiting one of the two different kinds of degeneracies. We show that $\dist$ is an upper bound for the spectral response strength~$\rca$ associated with the EP; see Fig.~\ref{fig:sketchdEPDP}. The important conclusion is that EPs generated by a small perturbation of a DP necessarily exhibit only a weak spectral response to perturbations, which renders a number of applications difficult.

The outline of the paper is as follows. Section~\ref{sec:MP} provides some mathematical preliminaries that are needed for the subsequent sections. Section~\ref{sec:dHDP} introduces the notion of the distance between a given Hamiltonian and the set of DPs. In Sec.~\ref{sec:srsEP} the relation to the spectral response strength is explored. Section~\ref{sec:passive} investigates the relevant aspects for passive systems. Various examples are considered in Sec.~\ref{sec:examples}. A summary is given in Sec.~\ref{sec:summary}.

\section{Matrix norms and distances}
\label{sec:MP}
The aim of this section is to review some aspects of linear algebra required in this paper. 

\subsection{Matrix norms}
\label{sec:norms}
A matrix norm $\normM{\op{A}}$ is defined as a mapping of a matrix $\op{A}$ to the nonnegative real numbers, see, e.g., Ref.~\cite{HJ13}, with the properties
\begin{eqnarray}
  && \normM{\op{A}+\op{B}}\leq\normM{\op{A}}+\normM{\op{B}} \;\text{(triangle inequality)}\ ,\\
  && \normM{\alpha \op{A}} = |\alpha|\,\normM{\op{A}} \; \text{(absolutely homogeneous)}\ ,\\
  && \normM{\op{A}} = 0 \,\text{ if and only if}\, \op{A}=0 \;\text{(definite)}\ ,\\
  \label{eq:MNsubm}
  && \normM{\op{A}\op{B}} \leq \normM{\op{A}}\,\normM{\op{B}} \;\text{(submultiplicative)} \ ,
\end{eqnarray}
for all matrices $\op{A}$ and $\op{B}$ and $\alpha\in\C$. The nonnegativity of $\normM{\cdot}$ follows from the first two items.

Important examples of matrix norms are the Frobenius norm
\begin{equation}\label{eq:Fronorm}
\normFro{\op{A}} := \sqrt{\trace{(\op{A}^\dagger\op{A})}} 
\end{equation}
with the trace $\trace$ and the spectral norm
\begin{equation}\label{eq:defspn}
\normspec{\op{A}} := \max_{\normV{\psi} = 1}\normV{\op{A}\psi} \ .
\end{equation}
We adopt the common but slightly confusing notation $\normspec{\cdot}$ both for the spectral matrix norm [in the left-hand side (LHS) of Eq.~(\ref{eq:defspn})] and the vector 2-norm $\normV{\psi} = \sqrt{\braket{\psi}{\psi}}$ of a vector $\ket{\psi}$ based on the usual inner product in complex vector space [in the right-hand side (RHS) of Eq.~(\ref{eq:defspn})].
Both matrix norms~(\ref{eq:Fronorm}) and~(\ref{eq:defspn}) share the important property of unitary invariance, i.e., 
\begin{equation}\label{eq:unitary}
\normM{\op{U}\op{A}\op{V}} = \normM{\op{A}} 
\end{equation}
for all matrices $\op{A}$ and all unitary matrices~$\op{U}$ and~$\op{V}$.
Moreover, both matrix norms~(\ref{eq:Fronorm}) and~(\ref{eq:defspn}) are compatible with the vector 2-norm, i.e., 
\begin{equation}\label{eq:compatible}
\normV{\op{A}\psi} \leq \normM{\op{A}}\,\normV{\psi} 
\end{equation}
for all matrices $\op{A}$ and vectors $\ket{\psi}$.
The following inequality holds for all matrices $\op{A}$
\begin{equation}\label{eq:sleqF}
\normspec{\op{A}} \leq \normFro{\op{A}} \ ,
\end{equation}
with equality for rank-1 matrices.

The calculation of the Frobenius norm~(\ref{eq:Fronorm}) is particularly easy
\begin{equation}\label{eq:FronormAij}
\normFro{\op{A}} = \sqrt{\sum_{ij}|A_{ij}|^2}
\end{equation}
where $A_{ij}$ are the matrix elements of $\op{A}$ in any orthonormal basis.
Also beneficial for us is that the Frobenius norm can be derived from an inner product,
\begin{equation}\label{eq:Fronormprod}
\normFro{\op{A}}^2 = \prodFro{\op{A}}{\op{A}} \ .
\end{equation}
The inner product is the Frobenius inner product, see, e.g., Ref.~\cite{HJ13}, of two matrices~$\op{A}$ and $\op{B}$,
\begin{equation}\label{eq:Fro}
\prodFro{\op{A}}{\op{B}} := \trace{(\op{A}^\dagger\op{B})} \ .
\end{equation}

\subsection{Matrix distances}
\label{sec:Mdist}
The distance between any two matrices $\op{A}$ and $\op{B}$ can be measured with the distance function (see, e.g., Ref.~\cite{Gentle17})
\begin{equation}\label{eq:df}
\distance{\op{A}}{\op{B}} := \normM{\op{A}-\op{B}} \ ,
\end{equation}
with arbitrary matrix norm. It is easy to show that this distance function fulfills the usual axioms of a metric:
\begin{eqnarray}
  \nonumber
  && \distance{\op{A}}{\op{B}} = 0 \,\text{if and only if}\, \op{A} = \op{B} \;\text{(identity of}\\
  && \hspace{1.88cm}\text{indiscernibles)}\\
  && \distance{\op{A}}{\op{B}} = \distance{\op{B}}{\op{A}} \;\text{(symmetry)}\\
  && \distance{\op{A}}{\op{B}} \leq \distance{\op{A}}{\op{C}}+\distance{\op{C}}{\op{B}} \;\text{(triangle inequality)}\hspace{0.6cm}
\end{eqnarray}
for all matrices~$\op{A}$, $\op{B}$, and $\op{C}$. From these axioms one can deduce $\distance{\op{A}}{\op{B}} \geq 0$.

It is important to mention that the assumed unitary invariance of the matrix norm~(\ref{eq:unitary}) transfers to the distance function~(\ref{eq:df}).

\section{Distance between a given Hamiltonian and the set of DPs}
\label{sec:dHDP}
In the mathematical literature several matrix nearness problems had been studied, as reviewed in Ref.~\cite{Higham89}. For example, one asks the question of how close a given nonnormal matrix~$\op{A}$ is to the set of normal matrices and which normal matrix~$\op{B}$ minimizes the distance function such as in Eq.~(\ref{eq:df}). A rather difficult task is to find the nearest matrix having at least two equal eigenvalues; see, e.g., Ref.~\cite{Demmel87}. Our problem is related but simpler. 

We consider an in general nonnormal $\order\times\order$-matrix $\op{H}$ and a non-Hermitian but normal $\order\times\order$-matrix $\HDP = \evDP\ident$ with a DP of order $\order$ with complex-valued eigenvalue $\evDP$ and $\order\times\order$-identity matrix $\ident$. Note that $\HDP$ is in the above form in any orthonormal basis.
Using the distance function~(\ref{eq:df}) we define the distance between $\op{H}$ and a given DP by $\distance{\op{H}}{\evDP\ident}$.
The distance between $\op{H}$ and the set of DPs of order $\order$ is then given by
\begin{equation}\label{eq:defdist}
  \dist(\op{H}) := \min\{\distance{\op{H}}{\evDP\ident}, \evDP\in\C\} \ .
\end{equation}

Introducing the traceless part of the Hamiltonian
\begin{equation}\label{eq:Htl}
\op{H}' := \op{H}-\bar{E}\ident 
\end{equation}
with the mean energy $\bar{E} := \trace{\op{H}}/\order\in\C$, we write
\begin{equation}
  \distance{\op{H}}{\evDP\ident} = \normM{\op{H}-\evDP\ident} = \normM{\op{H}'-(\evDP-\bar{E})\ident} \ .
\end{equation}
We now choose the Frobenius norm and evaluate with Eq.~(\ref{eq:Fronormprod})
\begin{equation}
  \normFro{\op{H}'-(\evDP-\bar{E})\ident}^2 = \normFro{\op{H}'}^2 + \order|\evDP-\bar{E}|^2 \ ,
\end{equation}
where we have exploited $\prodFro{\ident}{\ident} = \order$, $\prodFro{\ident}{\op{H}'} = \trace{\op{H}'} = 0$, and correspondingly $\prodFro{\op{H}'}{\ident} = 0$.  Hence we get
\begin{equation}
  \distance{\op{H}}{\evDP\ident} = \sqrt{\normFro{\op{H}'}^2 + \order|\evDP-\bar{E}|^2} \ .
\end{equation}
Clearly, this expression is minimal for $\evDP = \bar{E}$ and the minimizing $\HDP = \evDP\ident$ is unique. Plugging this into Eq.~(\ref{eq:defdist}) gives the first result
\begin{equation}\label{eq:distresult}
\dist(\op{H}) = \normFro{\op{H}'} \ .
\end{equation}
As the Frobenius norm is unitarily invariant, we can evaluate the matrix norm in Eq.~(\ref{eq:distresult}) in an orthonormal basis of our choice. 

From Eq.~(\ref{eq:FronormAij}) we can infer that the physical interpretation of the distance~(\ref{eq:distresult}) is that of an Euclidean distance defined on an $\order\times\order$ complex matrix space.
The matrix space is usually of higher dimension than the parameter space, i.e., the space spanned by the physical parameters relevant for a given system. However, this disadvantage is outweighed by the geometric tools provided by linear algebra, such as norms and inner products.

A small distance $\dist(\op{H})$ means a minor change of the matrix elements of $\op{H}$ compared with $\HDP$. This translates to a small detuning of the parameters of the system away from the DP.

\subsection{Hamiltonian without an EP}
For a nonnormal Hamiltonian~$\op{H}$ that does not have an EP, one can express the distance to the set of DPs in terms of the biorthogonal basis of the Hamiltonian
\begin{equation}\label{eq:biorthogonal}
\op{H}\ket{R_j} = E_j\ket{R_j}
\;\;\text{and}\;\;
\bra{L_j}\op{H} = E_j\bra{L_j}
\end{equation}
with the right eigenstates $\ket{R_j}$ and the left eigenstates $\ket{L_j}$; see, e.g., Ref.~\cite{CM98}. With $\braket{L_j}{R_l} = 0$ if $j\neq l$ and the normalization~$\braket{L_j}{R_j} = 1$ for all $j$, the biorthogonal expansion
\begin{equation}\label{eq:biorthoE}
\op{H} = \sum_j E_j\ket{R_j}\bra{L_j} \ ,
\end{equation}
and $\trace{(\ket{R_j}\bra{L_l})} = \braket{L_l}{R_j}$ it is straightforward to show with Eq.~(\ref{eq:distresult}) that
\begin{equation}\label{eq:dnnH}
\dist(\op{H}) = \sqrt{\sum_{j,l} (E_j^*-\bar{E}^*)(E_l-\bar{E})O_{lj}}
\end{equation}
with the $\order\times\order$ matrix 
\begin{equation}\label{eq:O}
O_{lj} := \braket{R_j}{R_l}\braket{L_l}{L_j} \ .
\end{equation}
This matrix is known as the nonorthogonality overlap matrix~\cite{CM98,FM02,Wiersig19}.
Its diagonal elements are the Petermann factors of the eigenstates~\cite{Schomerus09,HS17}, a measure of nonorthogonality. In Ref.~\cite{WLY20} the expression $\normFro{\op{H}'}$ (without relating it to a distance function) has been used to calculate the Petermann factors for the special case of $\order = 2$.

From Eq.~(\ref{eq:dnnH}), we learn that the distance of the Hamiltonian~$\op{H}$ to its nearest DP depends on the distance of its eigenvalues to the DP eigenvalue $\evDP = \bar{E}$ and the nonorthogonality of its eigenstates.
To emphasize, we do not advocate the use of Eq.~(\ref{eq:dnnH}) for calculating~$\dist(\op{H})$. It is much easier to directly evaluate Eq.~(\ref{eq:distresult}). Moreover, Eq.~(\ref{eq:dnnH}) is not valid for $\op{H}$ having an EP. At the EP the Petermann factors diverge and the expansion~(\ref{eq:biorthoE}) is not valid.

\subsection{Hamiltonian with an EP}
Next we consider the more interesting case of an $\order\times\order$-Hamiltonian $\op{H} = \HEP$ having an EP of order $\order$ with complex-valued eigenvalue $\evEP$ (frequency $\freqEP$ for optical systems). It follows from $\bar{E} = \evEP$ that the traceless part of the Hamiltonian [Eq.~(\ref{eq:Htl})] equals the operator
\begin{equation}\label{eq:N}
\op{N} := \HEP-\evEP\ident \ .
\end{equation}
The matrix $\op{N}$ is nilpotent of index $\order$, i.e., $\op{N}^\order = 0$ but $\op{N}^{\order-1} \neq 0$; see Refs.~\cite{Kato66,TE05,Wiersig22}. From Eq.~(\ref{eq:distresult}) it follows for an $\order\times\order$-Hamiltonian with an EP$_\order$ the important result
\begin{equation}\label{eq:distresultN}
\dist(\HEP) = \normFro{\op{N}} \ .
\end{equation}

The distance of an EP to the set of DPs in Eq.~(\ref{eq:distresultN}) can be related to Henrici's departure from normality~\cite{Henrici62}
\begin{equation} 
\Henrici(\op{A}) := \sqrt{\normFro{\op{A}}^2-\sum_{j=1}^\order |\lambda_j(\op{A})|^2}
\end{equation}
where $\lambda_j(\op{A})$ are the eigenvalues of the matrix $\op{A}$. Clearly, $\Henrici(\op{A}) \neq 0$ only if $\op{A}$ is nonnormal. In our case $\op{A} = \op{N}$ from Eq.~(\ref{eq:N}) has only zero eigenvalues as the matrix is nilpotent. As a consequence, the distance of a given EP to the set of DPs in Eq.~(\ref{eq:distresultN}) can be expressed by the departure of $\op{N}$ from normality,
\begin{equation} 
\dist(\HEP) = \Henrici(\op{N}) \ .
\end{equation}
It is known that Henrici's departure from normality is an upper bound for the distance of a matrix to the set of normal matrices~\cite{Higham89}. The quantity $\dist(\HEP)$ is therefore also an upper bound of the distance of the matrix $\HEP$ to the set of normal matrices.

From the physics perspective we can consider the departure of $\op{N}$ from normality and the distance of an EP to the set of DPs in many cases as a vague measure of the experimental effort in realizing the EP when starting at or near the minimizing DP. We come back to this point later when discussing some examples in Sec.~\ref{sec:examples}.

\section{Relation to the spectral response strength}
\label{sec:srsEP}
In Ref.~\cite{Wiersig22} the spectral response strength of a system with an EP of order~$\order$ has been determined to be 
\begin{equation}\label{eq:rca}
\rca = \normspec{\op{N}^{\order-1}} = \normFro{\op{N}^{\order-1}} \ .
\end{equation}
The spectral and Frobenius norms give in this particular case the same result since $\op{N}^{\order-1}$ has rank 1, which is a consequence of the nilpotency of $\op{N}$.
The spectral response strength is unitarily invariant. It describes the response of the system to perturbations,
\begin{equation}\label{eq:H}
\Ha = \Hs+\varepsilon\Hp 
\end{equation}
with $\Hs = \HEP$. The spectral response strength shows up as a factor in the bound of the eigenvalue splittings 
\begin{equation}\label{eq:specresponse}
|\ev_j-\evEP|^\order \leq \varepsilon \normspec{\Hp}\,\rca  \ ,
\end{equation}
where higher orders in the perturbation strength~$\varepsilon$ are ignored. The scaling of the splittings $|\ev_j-\evEP|$ with the $\order$th root of~$\varepsilon$ expresses the enhanced sensitivity of the EP with respect to perturbations.
It is important to mention that $\rca$ also bounds the intensity response to excitations~\cite{Wiersig22}:
\begin{equation}\label{eq:rtoe}
\normV{\psi}^{\text{EP}} \leq P\, \frac{1}{|\hbar\pf-\evEP|^{\order}}\, \rca\ ,
\end{equation}
where the vector $\ket{\psi}\propto e^{-i\pf t}$ is the long-time asymptotics of the equation of motion 
\begin{equation}\label{eq:iSe}
i\hbar\frac{d}{dt}\ket{\psi} = \Hs\ket{\psi} + e^{-i\pf t}P\ket{p}
\end{equation}
with the excitation power $P\geq 0$, the excitation frequency $\pf\in\R$, and a generic excitation vector $\ket{p}$ normalized to unity.

Comparing Eq.~(\ref{eq:distresultN}) with Eq.~(\ref{eq:rca}) reveals that, in the special case of an EP$_2$:
\begin{equation}\label{eq:rcadelta2}
\rca = \dist(\HEP) \;\text{for}\; \order = 2 \ .
\end{equation}
Hence the spectral response strength associated with the EP$_2$ and its distance to the set of DPs in the Frobenius norm is the same. This is different for higher-order EPs.
With the submultiplicativity~(\ref{eq:MNsubm}), we could use
\begin{equation}\label{eq:boundNsm}
\normFro{\op{N}^{\order-1}} \leq \normFro{\op{N}}^{\order-1} 
\end{equation}
to derive an upper bound for $\rca$ determined by $\dist(\HEP)$. However, a more stringent bound can be obtained by taking advantage of the nilpotency of $\op{N}$. For a nilpotent matrix $\op{N}$ with index $\order > 2$ holds~\cite{Gil03}
\begin{equation}
\normFro{\op{N}^k} \leq \gamma_{\order,k}\normFro{\op{N}}^k 
\end{equation}
with $k = 1,\ldots, n-1$ and positive numbers $\gamma_{\order,k}$ given by
\begin{equation}
  \gamma^2_{\order,k} = \frac{(n-1)(n-2)\cdot\ldots\cdot (n-k)}{(n-1)^kk!} \ .
\end{equation}
For $k=n-1$ we get
\begin{equation}\label{eq:boundNnp}
\normFro{\op{N}^{\order-1}} \leq (n-1)^{-\frac{n-1}{2}}\normFro{\op{N}}^{\order-1} \ .
\end{equation}
This inequality, valid for $\op{N}$ being nilpotent of index $\order > 2$, provides a much sharper bound than the inequality~(\ref{eq:boundNsm}). Combining the inequality~(\ref{eq:boundNnp}) with Eqs.~(\ref{eq:distresultN}) and~(\ref{eq:rca}) we arrive at our next crucial result
\begin{equation}\label{eq:rcadelta}
\rca \leq \left(\frac{\dist(\HEP)}{\sqrt{n-1}}\right)^{\order-1} \ .
\end{equation}
The spectral response strength associated with the EP is therefore bounded by its distance to the set of DPs in the Frobenius norm. The important implication is: if the EP is generated by a small perturbation of a DP then the spectral response strength of the resulting EP is weak. Clearly, this has a profound impact on applications such as EP-based sensing.
It is important to mention that this statement remains true if the DP which is (experimentally) used to generate the EP is not the minimizing DP. A non-minimizing DP has a larger distance $\dist' > \dist$ to the EP but the inequality~(\ref{eq:rcadelta}) is clearly also valid with $\dist$ replaced by~$\dist'$.

Note that, here, small perturbation means small compared with a perturbation needed to create another EP (of the same order) out of a DP. That one has a potentially larger spectral response strength than the first EP, as illustrated in Fig.~\ref{fig:sketchdEPDP}.

Another implication is that an EP which, under parameter variation, approaches a DP, has a vanishing response strength. In this sense a DP is an EP with zero response strength.

Equation~(\ref{eq:rcadelta}) can also be seen as an estimate for the spectral response strength~$\rca$. The calculation of $\dist(\HEP)$ is considerably simpler and faster.

\begin{figure}[ht]
\includegraphics[width=0.75\columnwidth]{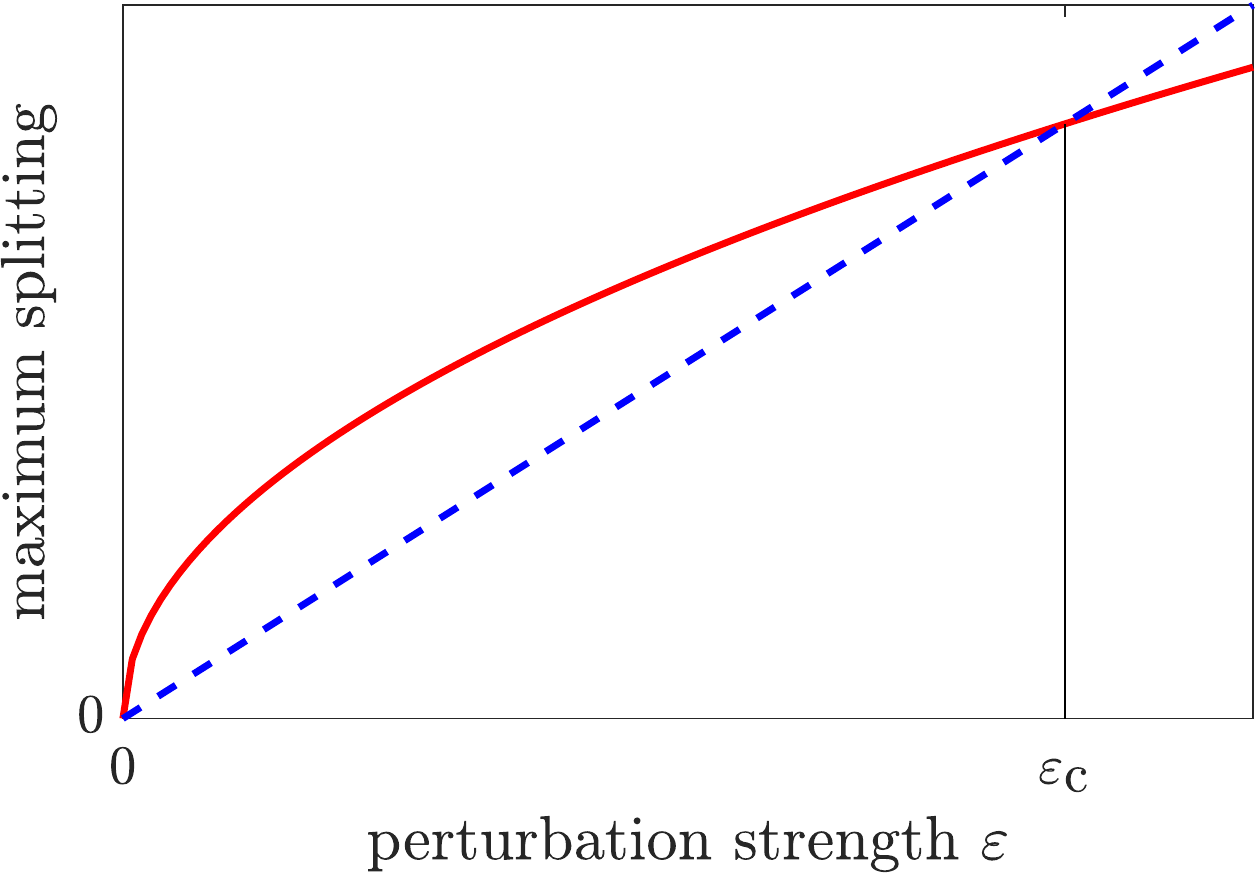}
\caption{Illustration of a comparison of the maximum energy (frequency) splitting for an EP$_2$ (red solid curve) according to the inequality~(\ref{eq:specresponse}) with its nearest DP$_2$ (blue dashed line) according to the inequality~(\ref{eq:specresponseDP}) under a perturbation. The vertical line marks the critical perturbation strength $\varepsilon_\text{c}$ below which the maximum splitting of the EP is larger.}
\label{fig:EPDPcom}
\end{figure}

In the context of the response strength of a system with an EP the physical interpretation of the nearest DP is that of a reference. To discuss this, consider the spectral response to a perturbation~$\varepsilon\Hp$. For EP-based sensors it is common to compare with a related DP which is chosen rather ad hoc by removing coupling terms in the Hamiltonian; see, e.g., Ref.~\cite{Wiersig14b}. In contrast to the inequality~(\ref{eq:specresponse}) the response of a system with a DP to a perturbation is (see the Appendix for a derivation)
\begin{equation}\label{eq:specresponseDP}
|\ev_j-\evDP| \leq \varepsilon \normspec{\Hp} \ .
\end{equation}
With the condition $\evDP = \evEP$ it is easy to show that the maximum splitting of the DP in inequality~(\ref{eq:specresponseDP}) equals the maximum splitting of the EP in inequality~(\ref{eq:specresponse}), cf. Fig.~\ref{fig:EPDPcom}, if
\begin{equation}
\varepsilon_\text{c}\normspec{\Hp} = \rca^{1/(\order-1)} \ ,
\end{equation}
assuming that the perturbation is still small enough such that the inequality~(\ref{eq:specresponse}) is valid. This elementary calculation reveals that the critical perturbation strength~$\varepsilon_\text{c}$ below which the maximum energy splitting related to the EP exceeds that of the reference DP (which is the nearest DP in matrix space here) is determined by $\rca$. 

It is mentioned that for the special case of $\order = 2$, the distance $\dist(\HEP)$ can also be related to the eigenstate response strength~$\rcb$ defined in Ref.~\cite{Wiersig22}. It measures the perturbation-induced change of the eigenvectors with respect to the EP eigenvector. For an EP$_2$ the simple relation $\rcb = 1/\rca$ holds. With Eq.~(\ref{eq:rcadelta2}) follows $\rcb = 1/\dist(\HEP)$. No such relation is known for $\order > 2$.

\section{Passive systems}
\label{sec:passive}
In Ref.~\cite{Wiersig22} it had been shown for the special cases $\order = 2$ and $\order = 3$ that there is an upper bound for the spectral response strength~$\rca$ in passive (no gain) systems. This upper bound limits the applicability of the given EP in the sensing setting.
In this section we derive an upper bound for the distance~$\dist$ for arbitrary $\order$. With the inequality~(\ref{eq:rcadelta}) this in turn generalizes the upper bound for $\rca$ in Ref.~\cite{Wiersig22}.

For passive systems the Hermitian decay operator
\begin{equation}\label{eq:decayop}
\oHami := i(\Hs-\Hs^\dagger) 
\end{equation}
is positive semidefinite; see, e.g., Refs.~\cite{Wiersig16,Wiersig19}. With Eq.~(\ref{eq:N}) we write 
\begin{equation}\label{eq:oHamiNN}
\op{N}-\op{N}^\dagger = -i(\oHami-\beta\ident) 
\end{equation}
where we have introduced the real number
\begin{equation}\label{eq:beta}
\beta := -2\imagc{\evEP} \ .
\end{equation}
Considering the fact that both $\op{N}$ and $\op{N}^\dagger$ are traceless because of their nilpotency~\cite{HJ13} one gets
\begin{equation}\label{eq:troHami}
\trace{\oHami} = \order\beta \ .
\end{equation}
For a positive semidefinite matrix $\trace{\oHami}\geq 0$ and hence $\beta \geq 0$.
Taking the Frobenius norm on both sides of Eq.~(\ref{eq:oHamiNN}) gives
\begin{equation}\label{eq:oHamiNNnorm}
\normFro{\op{N}-\op{N}^\dagger} = \normFro{\oHami-\beta\ident} \ . 
\end{equation}
In the following we assume that $\op{N}$ is a strictly upper triangular matrix. This can always be achieved by a unitary transformation of the Hamiltonian~$\HEP$ which does not change the Frobenius norm; see Eq.~(\ref{eq:unitary}). According to the Schur theorem (see, e.g., Ref.~\cite{Gil03}), a unitary transformation exists that transforms the matrix $\op{N}$ defined in Eq.~(\ref{eq:N}) to the sum of a diagonal matrix and a strictly upper triangular matrix. The former matrix is identical to zero because of the nilpotency of $\op{N}$. Obviously, if $\op{N}$ is a strictly upper triangular matrix then $\op{N}^\dagger$ is a strictly lower triangular matrix. With Eq.~(\ref{eq:FronormAij}) it follows for the LHS of Eq.~(\ref{eq:oHamiNNnorm})
\begin{equation}
\normFro{\op{N}-\op{N}^\dagger} = \sqrt{2}\normFro{\op{N}} \ . 
\end{equation}
With Eq.~(\ref{eq:Fronorm}) and the Hermiticity of $\oHami$ we can rewrite Eq.~(\ref{eq:oHamiNNnorm}) then as
\begin{equation}\label{eq:NN2}
\normFro{\op{N}} = \frac{1}{\sqrt{2}}\sqrt{\trace{\left[(\oHami-\beta\ident)^2\right]}} \ . 
\end{equation}
The positive semidefiniteness of~$\oHami$ restricts the maximum value of the RHS of Eq.~(\ref{eq:NN2}). It is attained when $\oHami$ is a rank-1 matrix. In this case, $\oHami$ has only one nonzero eigenvalue, which, according to Eq.~(\ref{eq:troHami}), must be $\order\beta$. With Eqs.~(\ref{eq:distresultN}) and~(\ref{eq:beta}) the important result follows:
\begin{equation}\label{eq:NN3}
\dist(\HEP) \leq \sqrt{2n(n-1)}|\imagc{\evEP}| \ . 
\end{equation}
This inequality valid for passive systems shows that the distance of a given EP to the set of DPs is bounded essentially by the decay rate at the EP. This bound can be transferred to the spectral response strength via the inequality~(\ref{eq:rcadelta}). It follows the next important result
\begin{equation}\label{eq:rcapassve}
\rca \leq \left(\sqrt{2n}|\imagc{\evEP}|\right)^{\order-1} \ .
\end{equation}
For $\order = 2$ the bound in Ref.~\cite{Wiersig22} for passive systems is recovered. For $\order = 3$ a slightly lower bound is obtained here. This improved bound is consistent with the numerical data in Ref.~\cite{Wiersig22}. To emphasis, Eq.~(\ref{eq:rcapassve}) is valid for all~$\order\geq 2$. It is therefore a generalization of the results proven in Ref.~\cite{Wiersig22}.

Note that for small perturbation strength~$\varepsilon$, the inequality~(\ref{eq:rcapassve}) together with the bound of the eigenvalue splittings~(\ref{eq:specresponse}) provide a much stronger statement than $\imagc{\ev_j} \leq 0$ which must also be fulfilled in passive systems.

The importance of the inequality~(\ref{eq:rcapassve}) becomes clear by remarking that $2|\imagc{\evEP}|$ is the linewidth of the spectral peak at the EP. Hence the upper bound for $\rca$ in the inequality~(\ref{eq:rcapassve}) limits the resolvability of the frequency splittings under perturbation and therefore the performance of a sensor based on such an EP.

The result in the inequality~(\ref{eq:rcapassve}) also has significant consequences for the intensity response of passive systems to excitations. Consider the DP-analog of inequality~(\ref{eq:rtoe}) (see the Appendix for a short derivation), 
\begin{equation}\label{eq:DPrtoe}
\normV{\psi}^{\text{DP}} = P\, \frac{1}{|\hbar\pf-\evDP|} \ .
\end{equation}
We compare the response of the two kinds of degeneracies at resonance, i.e., at the EP [inequality~(\ref{eq:rtoe})] with $\hbar\pf = \real{\evEP}$ and at the DP [Eq.~(\ref{eq:DPrtoe})] with $\hbar\pf = \real{\evDP}$. Using the inequality~(\ref{eq:rcapassve}) we obtain
\begin{equation}\label{eq:EPDPintensity}
\frac{\normV{\psi}^{\text{EP}}}{\normV{\psi}^{\text{DP}}} \leq (2\order)^{(n-1)/2} \ .
\end{equation}
This inequality tells us how much an EP$_\order$ can enhance the intensity response to excitation in a passive system if compared with a reference DP$_\order$. For $\order = 2$ the maximal enhancement factor is 2 which is consistent with the findings in Ref.~\cite{Sunada18}. For $\order = 3$ the maximal enhancement factor is 6.

\section{Examples}
\label{sec:examples}
In this section we provide several examples to illustrate our approach.

\subsection{Exceptional ring in wave-number space}
Our first example stresses the difference between the commonly used parameter space and wave-number space on the one side and the matrix space discussed in this paper on the other side.
In Ref.~\cite{ZHI15} it has been shown that EPs can be created out of a Dirac point in a photonic crystal slab. The finite thickness of the slab introduces radiation losses which are dissimilar for the two involved dipole and quadrupole modes. The effective Hamiltonian to describe this situation is
\begin{equation}\label{eq:Hring}
\op{H} = \left(\begin{array}{cc}
\omega_0     & \vg |\vc{k}| \\
\vg |\vc{k}| & \omega_0-i\gamma\\
\end{array}\right) \ ,
\end{equation}
where $\vc{k} = (k_x,k_y)$ is the two-dimensional wave vector, $\gamma \geq 0$ is the loss rate of the dipole mode (the loss of the quadrupole mode can be ignored), $\omega_0 > 0$ is the frequency of the unperturbed Dirac point, and $\vg > 0$ is the group velocity at the Dirac point. The eigenvalues of the Hamiltonian~(\ref{eq:Hring}) are
\begin{equation}\label{eq:oring}
\omega_\pm = \omega_0-i\frac{\gamma}{2}\pm\vg\sqrt{|\vc{k}|^2-\kc^2}
\end{equation}
with the critical wave number $\kc := \gamma/(2\vg)$. In the absence of radiation ($\gamma = 0$), Eq.~(\ref{eq:oring}) describes the linear dispersion near the Dirac point, $\omega_\pm = \omega_0\pm\vg|\vc{k}|$, the well-known Dirac cones. 

If the loss rate $\gamma$ is nonzero, even if it is very small, the system possesses an EP at $|\vc{k}| = \kc$. In the two-dimensional wave-number space $(k_x,k_y)$ with all other parameters fixed this degeneracy therefore appears as a ring of EPs. In the three-dimensional parameter space $(\omega_0, \vg |\vc{k}|, \gamma)$ this degeneracy is the two-dimensional plane $\vg |\vc{k}| = \gamma/2$. Both the parameter space and the wave-number space are clearly different from the $2\times 2$-dimensional complex matrix space where the distance $\dist(\HEP)$ of the EP to the set of DPs is defined. It is here calculated by inserting the Hamiltonian~(\ref{eq:Hring}) at the EP and its eigenvalue $\freqEP = \omega_0-i\frac{\gamma}{2}$ via Eq.~(\ref{eq:N}) into Eq.~(\ref{eq:distresultN})
\begin{equation}\label{eq:distring}
\dist(\HEP) = \gamma\ .
\end{equation}
This result makes intuitive sense because it is here that the loss turns the DP into an EP. A small loss rate $\gamma$ is sufficient but, according to Eq.~(\ref{eq:rcadelta2}), the spectral response strength $\rca = \gamma$ is then also small. To enlarge the spectral response to perturbations one has to increase the loss in the photonic crystal slab by reducing the thickness of the slab which obviously has a limit.

One remark is in order. The DP from which the experiments starts is $\HDPexp = \omega_0\ident$ according to Eq.~(\ref{eq:Hring}), in the absence of radiation, and for $\vc{k} = 0$ at a Dirac point. The minimizing DP is, however, $\HDP = \freqEP\ident = (\omega_0-i\frac{\gamma}{2})\ident$. The distance of $\HEP$ to $\HDPexp$ is $\sqrt{3/2}\gamma$ and therefore slightly larger than $\dist(\HEP)$, see Eq.~(\ref{eq:distring}). The above conclusions are still correct.

Note that $\gamma = 2|\imagc{\freqEP}|$, so that the inequality~(\ref{eq:rcapassve}) for passive systems holds here with the equal sign.

\subsection{Unidirectionally coupled pair of \PT-symmetric dimers}
Our second example is a higher-order EP at which $\dist$ and~$\rca$ are not equal. We consider a unidirectionally coupled pair of parity-time (\PT)-symmetric dimers introduced in the context of hierarchical construction of higher-order EPs~\cite{ZKO20}. The Hamiltonian is
\begin{equation}\label{eq:HcPT}
\op{H} = \left(\begin{array}{cccc}
\omega_0-i\alpha & g         & 0             & 0 \\
g           & \omega_0+i\alpha & 0           & 0 \\
\kappa      & 0           & \omega_0-i\alpha & g \\
0           & 0           & g             & \omega_0+i\alpha \\
\end{array}\right) \ .
\end{equation}
Each of the two $2\times 2$ subblocks along the diagonal describes a \PT-symmetric dimer with real-valued frequency~$\omega_0$, gain/loss coefficient $\alpha > 0$, and internal coupling coefficient $g > 0$. These two dimers are unidirectionally coupled with the strength $\kappa > 0$. 
This system can be realized experimentally by two evanescently coupled microrings each with two modes, one traveling clockwise and one counterclockwise~\cite{ZKO20}. One microring exhibits gain, the other one exhibits an equal amount of loss. The unidirectional coupling can be achieved by evanescently coupling the lossy microring to a semi-infinite waveguide with an end mirror~\cite{ZOE20}. This introduces a fully asymmetric backscattering~\cite{Wiersig18b} between the traveling waves in that microring. 

For $\alpha = g$ the Hamiltonian~(\ref{eq:HcPT}) possesses an EP$_4$ with real eigenvalue $\freqEP = \omega_0$. Inserting this eigenvalue and the Hamiltonian~(\ref{eq:HcPT}) via Eq.~(\ref{eq:N}) into Eq.~(\ref{eq:distresultN}) gives
\begin{equation}
\dist(\HEP) = \sqrt{8g^2+\kappa^2}\ .
\end{equation}
The Pythagoras-like appearance of $\dist(\HEP)$ in this case is obvious. The experimental effort in realizing a significant $\dist(\HEP)$ starting from uncoupled microrings/resonators is based on implementing either a large internal coupling $g$ or a large unidirectional coupling~$\kappa$. 
For the spectral response strength~(\ref{eq:rca}) we get
\begin{equation}\label{eq:rcaunid}
\rca = 2g^2\kappa \ .
\end{equation}
It can be easily verified that the inequality~(\ref{eq:rcadelta}) is satisfied. Equation~(\ref{eq:rcaunid}) tells us that large $g$ or large $\kappa$ is only a necessary but not a sufficient condition for getting a large response from the system at the EP. For the sufficient condition the product $g^2\kappa$ must be large.
The bound in the inequality~(\ref{eq:rcapassve}) does not apply since the system is not passive.

\subsection{Coupling of optical modes with different angular momenta}
The third example is a deformed microdisk cavity~\cite{ND97,GCNNSFSC98}. The broad range of applications of deformed microcavities is reviewed in Ref.~\cite{CW15}. For generating EPs in this kind of system a weak boundary deformation is sufficient~\cite{YKW18,KYW18}. Such a boundary deformation can be expressed in polar coordinates as $r(\phi) = R + f(\phi)$ with the radius of the unperturbed microdisk $R$ and the deformation function~$|f(\phi)|\ll R$. We consider here only deformation functions that preserve a mirror-reflection symmetry. Moreover, we restrict ourselves to two modes. The mode with lower radiation losses is called mode~1. Its complex frequency is $\omega_1$ and its azimuthal mode number is $m$. Mode~2 has the higher radiation losses, complex frequency $\omega_2$, and azimuthal mode number $p < m$. If the frequencies of the two modes are nearly degenerate, i.e., $\omega_1 \approx \omega_2$, a first-order perturbation theory can be applied leading to the effective non-Hermitian Hamiltonian~\cite{KYW18}
\begin{equation}\label{eq:Heff}
\op{H} = \left(\begin{array}{cc}
x_1 & 0 \\
0   & x_2\\
\end{array}\right) - 
x_1\left(\begin{array}{cc}
A_{mm}^{\text{e/o}} & A_{mp}^{\text{e/o}} \\
A_{pm}^{\text{e/o}} & A_{pp}^{\text{e/o}}\\
\end{array}\right) \ .
\end{equation}
The eigenvalues of this Hamiltonian are the frequencies of the two modes in the deformed microcavity. Here, $x_i = \omega_i R/c$ are the dimensionless complex frequencies of the unperturbed modes and $c$ is the speed of light in vacuum. The Fourier harmonics of the deformation function are given for the even and odd parity as
\begin{eqnarray}
\label{eq:Agenerale}
A_{pm}^{\text{e}} & = & \frac{\varepsilon_p}{\pi R}\int_0^\pi f(\phi)\cos{(p\phi)}\cos{(m\phi)}d\phi \ ,\\
\label{eq:Ageneralo}
A_{pm}^{\text{o}} & = & \frac{\varepsilon_p}{\pi R}\int_0^\pi f(\phi)\sin{(p\phi)}\sin{(m\phi)}d\phi \ ,
\end{eqnarray}
with $\varepsilon_p = 2$ if $p \neq 0$ and $\varepsilon_p = 1$ otherwise. Restricting to the relevant case $m,p > 0$ the matrix elements $A_{mp}^{\text{e/o}}$ and $A_{pm}^{\text{e/o}}$ are equal for fixed parity. 

Note that the unperturbed system, i.e., the undeformed microdisk, is strictly speaking not at an DP as $x_1\neq x_2$. However, the distance to a DP is very small, as $x_1\approx x_2$ has been required.

The Hamiltonian~(\ref{eq:Heff}) has an EP if
\begin{equation}
[x_1-x_2-x_1(A_{mm}^{\text{e/o}}-A_{pp}^{\text{e/o}})]^2+4x_1^2(A_{mp}^{\text{e/o}})^2 = 0 \ .
\end{equation}
The eigenvalue of the EP is 
\begin{equation}
x_{\text{EP}} = \frac{x_1+x_2}{2}-x_1\frac{A_{mm}^{\text{e/o}}+A_{pp}^{\text{e/o}}}{2} \ .
\end{equation}
Plugging this eigenvalue and the Hamiltonian~(\ref{eq:Heff}) by means of Eq.~(\ref{eq:N}) into Eq.~(\ref{eq:distresultN}) gives the distance to the set of DPs
\begin{equation}
\dist(\HEP) = 2|x_1 A_{mp}^{\text{e/o}}|\ .
\end{equation}
Clearly, a weak deformation ($|A_{mp}^{\text{e/o}}| \ll 1$) implies a small distance $\dist(\HEP)$. This is turn leads, according to Eq.~(\ref{eq:rcadelta2}), to a weak spectral response strength $\rca$ to perturbations (e.g. induced by a nanoparticle close to the boundary of the microcavity). The response can therefore be enhanced if the deformation is increased. This, however, often leads to enhanced radiation losses~\cite{Noeckel94}. Also note that if the deformation is too strong then the perturbation theory for weak boundary deformations is no longer valid.

\subsection{Fully asymmetric hopping model}
\label{sec:HNmodel}
Here, we study an $\order\times\order$ Hamiltonian describing uniform and unidirectional hoppings in a nearest-neighbor tight-binding chain 
\begin{equation}\label{eq:HNmodel}
\HEP = \left(\begin{array}{ccccc}
\evEP    & A    & 0       & \ldots & 0  \\
0      & \evEP    & A     & \ldots & 0  \\
0      & 0      & \evEP     & \ldots & 0  \\
\vdots & \vdots & \vdots  &  \ddots       & \vdots   \\
0      & 0 & 0       &  \ldots     & \evEP\\
\end{array}\right) \ .
\end{equation}
If the complex hopping parameter $A$ is nonzero, the Hamiltonian has an EP$_\order$ with eigenvalue $\evEP$. The Hamiltonian~(\ref{eq:HNmodel}) can be considered as the nonperiodic, fully asymmetric limiting case of the Hatano-Nelson model of a cylindrical superconductor~\cite{HN96}.  
The spectral response strength has been determined in Ref.~\cite{Wiersig22} to be
\begin{equation}\label{eq:HNmodelrca}
\rca = |A|^{\order-1} \ .
\end{equation}

The distance of this EP to the set of DPs can be easily calculated from Eqs.~(\ref{eq:distresultN}), (\ref{eq:N}), and~(\ref{eq:FronormAij}) yielding
\begin{equation} 
\dist(\HEP) = \sqrt{\order-1}|A| \ .
\end{equation}
With Eq.~(\ref{eq:HNmodelrca}) this gives
\begin{equation}
\rca = \left(\frac{\dist(\HEP)}{\sqrt{n-1}}\right)^{\order-1} \ .
\end{equation}
Interestingly, this is the inequality~(\ref{eq:rcadelta}) with the equality sign. Hence, this fully asymmetric hopping model has the largest possible spectral response strength $\rca$ for given distance $\dist(\HEP)$.

\subsection{Random Hamiltonians at EPs}
Finally, we adopt the random-matrix approach invented in Ref.~\cite{Wiersig22} to numerically generate a whole class of examples. To do so, we introduce the $\order\times\order$ matrix $\HEP$ having an EP$_\order$ with eigenvalue $\evEP$ via a similarity transformation $\HEP = \op{Q}\op{J}\op{Q}^{-1}$, with $\op{J}$ being an $\order\times\order$ matrix with an EP$_\order$ in Jordan normal form and $\op{Q}$ is an, in general nonunitary, $\order\times\order$ matrix consisting of complex random numbers with real and imaginary parts being drawn from a uniform distribution on the interval $[-1/2,1/2]$. Clearly, $\HEP$ is not completely random, but nevertheless we refer to it as the \quoting{random Hamiltonian with an EP}.

\begin{figure}[ht]
\includegraphics[width=0.95\columnwidth]{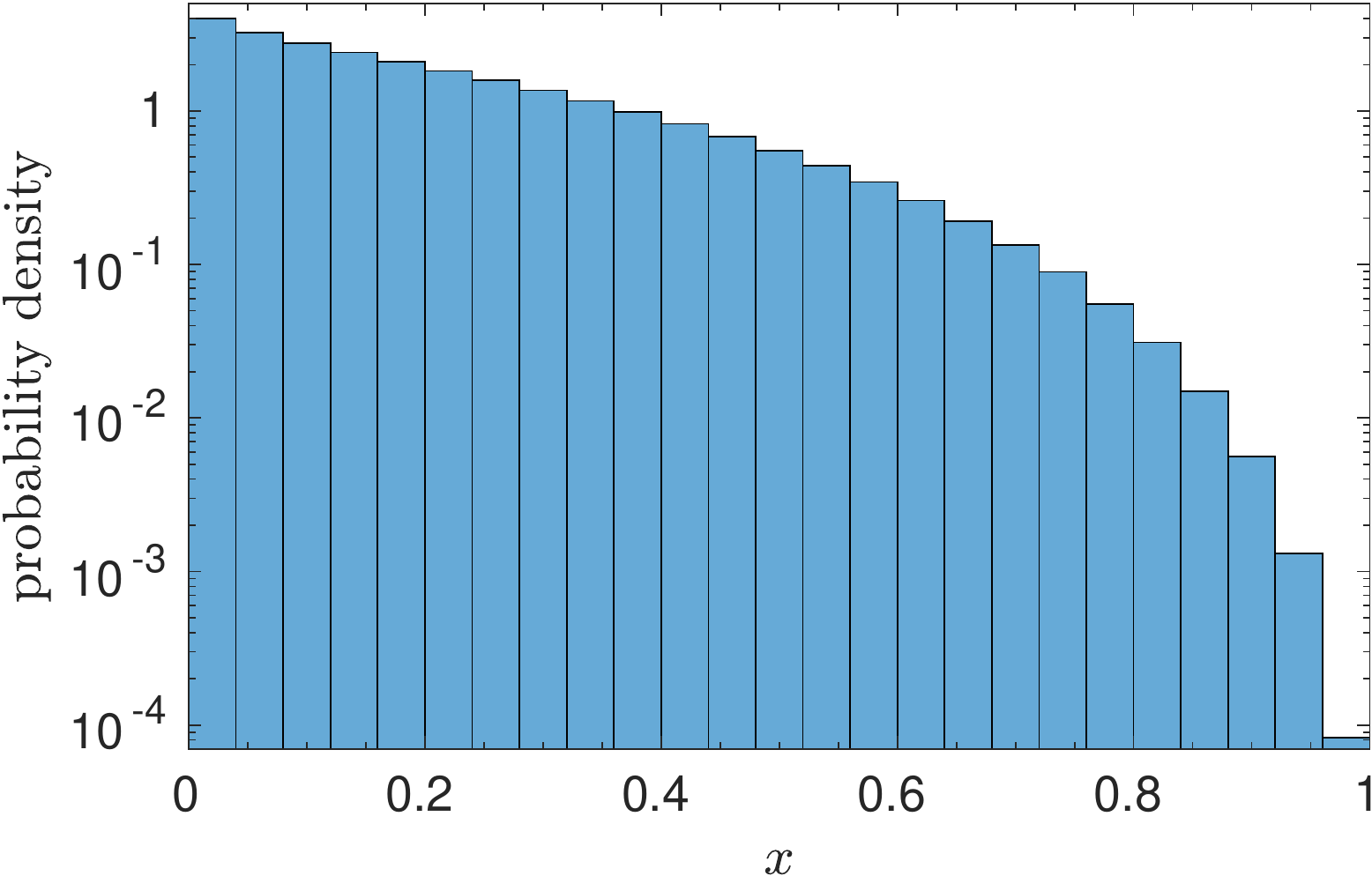}
\caption{Probability density function (notice the logarithmic scale) of the dimensionless number $x$ defined in Eq.~(\ref{eq:x1}) computed from $10^8$ realizations of random Hamiltonians having an EP$_4$ with eigenvalue $\evEP = -i0.5$.}
\label{fig:historcadelta}
\end{figure}

To numerically confirm the inequality~(\ref{eq:rcadelta}) we define the nonnegative quantity
\begin{equation}\label{eq:x1}
x := \frac{(n-1)^{(\order-1)/2}\rca}{\dist(\HEP)^{\order-1}} \ ,
\end{equation}
which should be less than or equal to unity. Figure~\ref{fig:historcadelta} shows a histogram resulting from $10^8$ realizations of random Hamiltonians for the case of an EP of order $\order = 4$. It can be clearly seen that $x \leq 1$ and therefore the upper bound for the response strength~$\rca$ given by the inequality~(\ref{eq:rcadelta}) is fulfilled. The average and the maximal value of $x$ are numerically determined to be $\approx 0.21$ and $\approx 0.99$. The latter value indicates that the upper bound given by the RHS of the inequality~(\ref{eq:rcadelta}) is sharp. We observe this numerical indication of a sharp upper bound also for $\order < 4$ but not for $\order > 4$. For instance, for $\order = 2,\ldots,6$ the maximal values of $x$ [Eq.~(\ref{eq:x1})] are $1$, $1$, $0.99$, $0.91$, and $0.74$. 

\begin{figure}[ht]
\includegraphics[width=0.95\columnwidth]{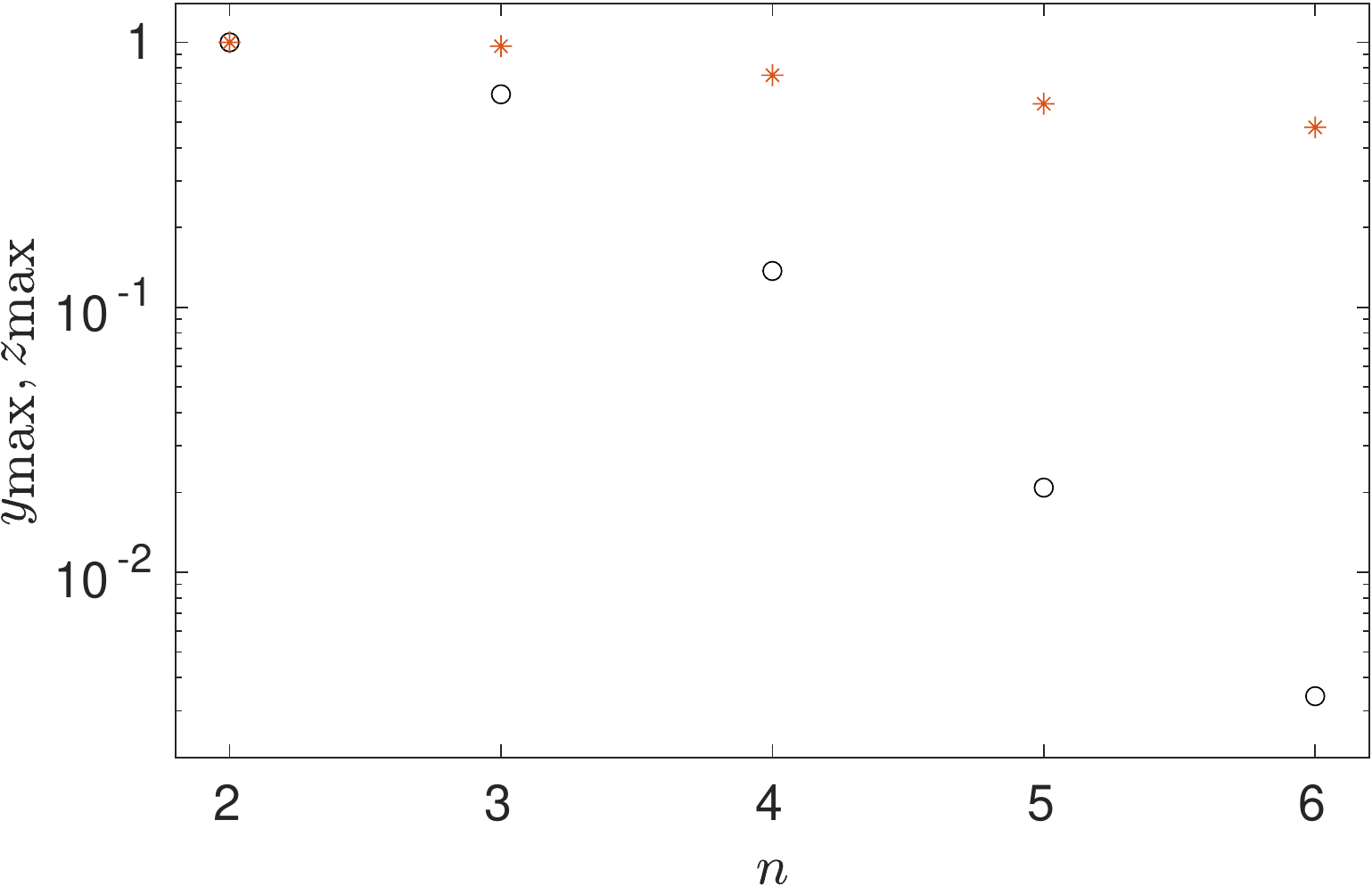}
\caption{Maximal values of the dimensionless distance $y$ [Eq.~(\ref{eq:x2}), star symbols] and the dimensionless response strength $z$ [Eq.~(\ref{eq:x3}), circle symbols] for passive systems as function of the order $\order$ of the EP. From $10^8$ realizations (for each $\order$) of random Hamiltonians having an EP$_\order$ with eigenvalue $\evEP = -i1.2$ those are selected that possess a positive semidefinite decay operator~$\oHami$ [Eq.~(\ref{eq:decayop})]. Note the logarithmic scale on the vertical axis.}
\label{fig:xDn}
\end{figure}

To check the upper bound for the distance $\dist$ [inequality~(\ref{eq:NN3})] and the spectral response strength~$\rca$ [inequality~(\ref{eq:rcapassve})] in passive systems, we consider the random Hamiltonians~$\HEP$ with an EP$_\order$ for several values of $\order$. We define the nonnegative quantities
\begin{equation}\label{eq:x2}
y := \frac{\dist(\HEP)}{\sqrt{2n(n-1)}|\imagc{\evEP}|} 
\end{equation}
and
\begin{equation}\label{eq:x3}
z := \frac{\rca}{\left(\sqrt{2n}|\imagc{\evEP}|\right)^{\order-1}} \ ,
\end{equation}
which according to the inequalities~(\ref{eq:NN3}) and~(\ref{eq:rcapassve}) are less than or equal to unity. 
For each value of $\order$, we select from $10^8$ realizations of random Hamiltonians those that have a positive semidefinite decay operator~$\oHami$ [Eq.~(\ref{eq:decayop})]. Figure~\ref{fig:xDn} shows the resulting maximal values of $y$ and $z$ versus the order $\order$. First, one can observe that $y_{\text{max}}, z_{\text{max}} \leq 1$. Hence the inequalities~(\ref{eq:NN3}) and~(\ref{eq:rcapassve}) are indeed fulfilled. Second, from $y_{\text{max}} \geq z_{\text{max}}$ we conclude that the inequality~(\ref{eq:NN3}) gives a sharper bound than the inequality~(\ref{eq:rcapassve}). This is understandable as the former bound results directly from the positive semidefiniteness of $\oHami$, whereas the latter additionally requires the inequality~(\ref{eq:rcadelta}).

\section{Summary}
\label{sec:summary}
We have introduced the concept of the distance $\dist$ of an $\order\times\order$ Hamiltonian to the set of DPs of order $\order$ in matrix space. This concept is mathematical in nature but it has physical relevance for non-Hermitian Hamiltonians with an EP. In this case, $\dist$ is a vague measure of the experimental effort to create an EP out of a DP of the same order.
Interestingly, $\dist$ determines an upper bound for the spectral response strength~$\rca$ of the given system with the EP. This not only provides an easy way to estimate $\rca$, but, more importantly, it also reveals that the strategy of experimentally implementing an EP by slightly modifying a DP results in an EP exhibiting only a weak response to perturbations.
We have related the distance $\dist$ to Henrici's departure from normality. 

For passive systems we have derived an upper bound for the distance~$\dist$. This bound engenders an upper bound for the spectral response strength $\rca$ which constitutes an improvement and generalization of the bound for $\rca$ in Ref.~\cite{Wiersig22} to EPs of arbitrary order.

Analytical and numerical results for various physical examples have been presented which demonstrated the new insights in the physics of EPs provided by the concept introduced in this paper.

\acknowledgments 
Valuable discussions with J. Kullig and R. El-Ganainy are acknowledged. 

\begin{appendix}
\section{Spectral and intensity response at a DP}
\label{app:srDP}
In this appendix we derive the spectral response to perturbations [inequality~(\ref{eq:specresponseDP})] and the intensity response to excitations [Eq.~(\ref{eq:DPrtoe})] for a system with a DP. To do so, we first consider the eigenvalue equation of the Hamiltonian~(\ref{eq:H}) 
\begin{equation}\label{eq:EP}
(\HDP+\varepsilon\Hp)\ket{\state_j} = \ev_j\ket{\state_j}
\end{equation}
with $\Hs = \HDP$, eigenvalues $\ev_j$ and eigenstates $\ket{\state_j}$.
With $\HDP = \evDP\ident$ we can write
\begin{equation}
\varepsilon\Hp\ket{\state_j} = (\ev_j-\evDP)\ket{\state_j} \ .
\end{equation}
Taking the vector 2-norm on both sides of this equation and using the normalization $\normV{\state_j} = 1$ gives
\begin{equation}
\varepsilon\normV{\Hp\state_j} = |\ev_j-\evDP| \ .
\end{equation}
Exploiting the compatibility of the matrix norm to the vector 2-norm~(\ref{eq:compatible}) and again the normalization of the eigenstate gives the inequality~(\ref{eq:specresponseDP}).

Next, we quickly derive the intensity response for $\op{H} = \HDP = \evDP\ident$ by solving Eq.~(\ref{eq:iSe}) with the ansatz $\ket{\psi}\propto e^{-i\pf t}$ to yield
\begin{equation}
\ket{\psi} = (E\ident-\evDP\ident)^{-1} e^{-i\pf t}P\ket{p}
\end{equation}
with $E = \hbar\pf$. Taking the vector 2-norm on both sides of this equation and using the normalization  $\normV{p} = 1$ gives Eq.~(\ref{eq:DPrtoe}).

\end{appendix}

\bibliography{../../bib/fg4,../../bib/extern}

\begin{thebibliography}{69}%
\makeatletter
\providecommand \@ifxundefined [1]{%
 \@ifx{#1\undefined}
}%
\providecommand \@ifnum [1]{%
 \ifnum #1\expandafter \@firstoftwo
 \else \expandafter \@secondoftwo
 \fi
}%
\providecommand \@ifx [1]{%
 \ifx #1\expandafter \@firstoftwo
 \else \expandafter \@secondoftwo
 \fi
}%
\providecommand \natexlab [1]{#1}%
\providecommand \enquote  [1]{``#1''}%
\providecommand \bibnamefont  [1]{#1}%
\providecommand \bibfnamefont [1]{#1}%
\providecommand \citenamefont [1]{#1}%
\providecommand \href@noop [0]{\@secondoftwo}%
\providecommand \href [0]{\begingroup \@sanitize@url \@href}%
\providecommand \@href[1]{\@@startlink{#1}\@@href}%
\providecommand \@@href[1]{\endgroup#1\@@endlink}%
\providecommand \@sanitize@url [0]{\catcode `\\12\catcode `\$12\catcode
  `\&12\catcode `\#12\catcode `\^12\catcode `\_12\catcode `\%12\relax}%
\providecommand \@@startlink[1]{}%
\providecommand \@@endlink[0]{}%
\providecommand \url  [0]{\begingroup\@sanitize@url \@url }%
\providecommand \@url [1]{\endgroup\@href {#1}{\urlprefix }}%
\providecommand \urlprefix  [0]{URL }%
\providecommand \Eprint [0]{\href }%
\providecommand \doibase [0]{https://doi.org/}%
\providecommand \selectlanguage [0]{\@gobble}%
\providecommand \bibinfo  [0]{\@secondoftwo}%
\providecommand \bibfield  [0]{\@secondoftwo}%
\providecommand \translation [1]{[#1]}%
\providecommand \BibitemOpen [0]{}%
\providecommand \bibitemStop [0]{}%
\providecommand \bibitemNoStop [0]{.\EOS\space}%
\providecommand \EOS [0]{\spacefactor3000\relax}%
\providecommand \BibitemShut  [1]{\csname bibitem#1\endcsname}%
\let\auto@bib@innerbib\@empty
\bibitem [{\citenamefont {Miri}\ and\ \citenamefont {Al\`{u}}(2019)}]{MA19}%
  \BibitemOpen
  \bibfield  {author} {\bibinfo {author} {\bibfnamefont {M.-A.}\ \bibnamefont
  {Miri}}\ and\ \bibinfo {author} {\bibfnamefont {A.}~\bibnamefont {Al\`{u}}},\
  }\bibfield  {title} {\bibinfo {title} {Exceptional points in optics and
  photonics},\ }\href {https://doi.org/10.1126/science.aar7709} {\bibfield
  {journal} {\bibinfo  {journal} {Science}\ }\textbf {\bibinfo {volume}
  {363}},\ \bibinfo {pages} {eaar7709} (\bibinfo {year} {2019})}\BibitemShut
  {NoStop}%
\bibitem [{\citenamefont {{\"O}zdemir}\ \emph {et~al.}(2019)\citenamefont
  {{\"O}zdemir}, \citenamefont {Rotter}, \citenamefont {Nori},\ and\
  \citenamefont {Yang}}]{ORN19}%
  \BibitemOpen
  \bibfield  {author} {\bibinfo {author} {\bibfnamefont {{\c{S}}.~K.}\
  \bibnamefont {{\"O}zdemir}}, \bibinfo {author} {\bibfnamefont
  {S.}~\bibnamefont {Rotter}}, \bibinfo {author} {\bibfnamefont
  {F.}~\bibnamefont {Nori}},\ and\ \bibinfo {author} {\bibfnamefont
  {L.}~\bibnamefont {Yang}},\ }\bibfield  {title} {\bibinfo {title}
  {Parity–time symmetry and exceptional points in photonics},\ }\href
  {https://doi.org/10.1038/s41563-019-0304-9} {\bibfield  {journal} {\bibinfo
  {journal} {Nat. Materials}\ }\textbf {\bibinfo {volume} {18}},\ \bibinfo
  {pages} {783} (\bibinfo {year} {2019})}\BibitemShut {NoStop}%
\bibitem [{\citenamefont {Kato}(1966)}]{Kato66}%
  \BibitemOpen
  \bibfield  {author} {\bibinfo {author} {\bibfnamefont {T.}~\bibnamefont
  {Kato}},\ }\href {https://doi.org/10.1007/978-3-642-66282-9} {\emph {\bibinfo
  {title} {Perturbation Theory for Linear Operators}}}\ (\bibinfo  {publisher}
  {Springer},\ \bibinfo {address} {New York},\ \bibinfo {year}
  {1966})\BibitemShut {NoStop}%
\bibitem [{\citenamefont {Heiss}(2000)}]{Heiss00}%
  \BibitemOpen
  \bibfield  {author} {\bibinfo {author} {\bibfnamefont {W.~D.}\ \bibnamefont
  {Heiss}},\ }\bibfield  {title} {\bibinfo {title} {Repulsion of resonance
  states and exceptional points},\ }\href
  {https://doi.org/10.1103/PhysRevE.61.929} {\bibfield  {journal} {\bibinfo
  {journal} {Phys. Rev. E}\ }\textbf {\bibinfo {volume} {61}},\ \bibinfo
  {pages} {929} (\bibinfo {year} {2000})}\BibitemShut {NoStop}%
\bibitem [{\citenamefont {Berry}(2004)}]{Berry04}%
  \BibitemOpen
  \bibfield  {author} {\bibinfo {author} {\bibfnamefont {M.~V.}\ \bibnamefont
  {Berry}},\ }\bibfield  {title} {\bibinfo {title} {Physics of nonhermitian
  degeneracies},\ }\href {https://doi.org/10.1023/B:CJOP.0000044002.05657.04}
  {\bibfield  {journal} {\bibinfo  {journal} {Czech. J. Phys.}\ }\textbf
  {\bibinfo {volume} {54}},\ \bibinfo {pages} {1039} (\bibinfo {year}
  {2004})}\BibitemShut {NoStop}%
\bibitem [{\citenamefont {Heiss}(2004)}]{Heiss04}%
  \BibitemOpen
  \bibfield  {author} {\bibinfo {author} {\bibfnamefont {W.~D.}\ \bibnamefont
  {Heiss}},\ }\bibfield  {title} {\bibinfo {title} {Exceptional points of
  non-{H}ermitian operators},\ }\href
  {https://iopscience.iop.org/article/10.1088/0305-4470/37/6/034} {\bibfield
  {journal} {\bibinfo  {journal} {J. Phys. A: Math. Gen.}\ }\textbf {\bibinfo
  {volume} {37}},\ \bibinfo {pages} {2455} (\bibinfo {year}
  {2004})}\BibitemShut {NoStop}%
\bibitem [{\citenamefont {Berry}\ and\ \citenamefont {Wilkinson}(1984)}]{BW84}%
  \BibitemOpen
  \bibfield  {author} {\bibinfo {author} {\bibfnamefont {M.~V.}\ \bibnamefont
  {Berry}}\ and\ \bibinfo {author} {\bibfnamefont {M.}~\bibnamefont
  {Wilkinson}},\ }\bibfield  {title} {\bibinfo {title} {Diabolic points in the
  spectra of triangles},\ }\href {https://doi.org/10.1098/rspa.1984.0022}
  {\bibfield  {journal} {\bibinfo  {journal} {Proc. R. Soc. Lond. A.}\ }\textbf
  {\bibinfo {volume} {392}},\ \bibinfo {pages} {15} (\bibinfo {year}
  {1984})}\BibitemShut {NoStop}%
\bibitem [{\citenamefont {Dembowski}\ \emph {et~al.}(2001)\citenamefont
  {Dembowski}, \citenamefont {Gr{\"a}f}, \citenamefont {Harney}, \citenamefont
  {Heine}, \citenamefont {Heiss}, \citenamefont {Rehfeld},\ and\ \citenamefont
  {Richter}}]{DGH01}%
  \BibitemOpen
  \bibfield  {author} {\bibinfo {author} {\bibfnamefont {C.}~\bibnamefont
  {Dembowski}}, \bibinfo {author} {\bibfnamefont {H.-D.}\ \bibnamefont
  {Gr{\"a}f}}, \bibinfo {author} {\bibfnamefont {H.~L.}\ \bibnamefont
  {Harney}}, \bibinfo {author} {\bibfnamefont {A.}~\bibnamefont {Heine}},
  \bibinfo {author} {\bibfnamefont {W.~D.}\ \bibnamefont {Heiss}}, \bibinfo
  {author} {\bibfnamefont {H.}~\bibnamefont {Rehfeld}},\ and\ \bibinfo {author}
  {\bibfnamefont {A.}~\bibnamefont {Richter}},\ }\bibfield  {title} {\bibinfo
  {title} {Experimental {O}bservation of the {T}opological {S}tructure of
  {E}xceptional {P}oints},\ }\href {https://doi.org/10.1103/PhysRevLett.86.787}
  {\bibfield  {journal} {\bibinfo  {journal} {Phys. Rev. Lett.}\ }\textbf
  {\bibinfo {volume} {86}},\ \bibinfo {pages} {787} (\bibinfo {year}
  {2001})}\BibitemShut {NoStop}%
\bibitem [{\citenamefont {Dembowski}\ \emph {et~al.}(2004)\citenamefont
  {Dembowski}, \citenamefont {Dietz}, \citenamefont {Gr{\"a}f}, \citenamefont
  {Harney}, \citenamefont {Heine}, \citenamefont {Heiss},\ and\ \citenamefont
  {Richter}}]{DDG04}%
  \BibitemOpen
  \bibfield  {author} {\bibinfo {author} {\bibfnamefont {C.}~\bibnamefont
  {Dembowski}}, \bibinfo {author} {\bibfnamefont {B.}~\bibnamefont {Dietz}},
  \bibinfo {author} {\bibfnamefont {H.-D.}\ \bibnamefont {Gr{\"a}f}}, \bibinfo
  {author} {\bibfnamefont {H.~L.}\ \bibnamefont {Harney}}, \bibinfo {author}
  {\bibfnamefont {A.}~\bibnamefont {Heine}}, \bibinfo {author} {\bibfnamefont
  {W.~D.}\ \bibnamefont {Heiss}},\ and\ \bibinfo {author} {\bibfnamefont
  {A.}~\bibnamefont {Richter}},\ }\bibfield  {title} {\bibinfo {title}
  {Encircling an exceptional point},\ }\href
  {https://doi.org/10.1103/PhysRevE.69.056216} {\bibfield  {journal} {\bibinfo
  {journal} {Phys. Rev. E}\ }\textbf {\bibinfo {volume} {69}},\ \bibinfo
  {pages} {056216} (\bibinfo {year} {2004})}\BibitemShut {NoStop}%
\bibitem [{\citenamefont {Dietz}\ \emph {et~al.}(2007)\citenamefont {Dietz},
  \citenamefont {Friedrich}, \citenamefont {Metz}, \citenamefont {Miski-Oglu},
  \citenamefont {Richter}, \citenamefont {Sch{\"a}fer},\ and\ \citenamefont
  {Stafford}}]{DFM07}%
  \BibitemOpen
  \bibfield  {author} {\bibinfo {author} {\bibfnamefont {B.}~\bibnamefont
  {Dietz}}, \bibinfo {author} {\bibfnamefont {T.}~\bibnamefont {Friedrich}},
  \bibinfo {author} {\bibfnamefont {J.}~\bibnamefont {Metz}}, \bibinfo {author}
  {\bibfnamefont {M.}~\bibnamefont {Miski-Oglu}}, \bibinfo {author}
  {\bibfnamefont {A.}~\bibnamefont {Richter}}, \bibinfo {author} {\bibfnamefont
  {F.}~\bibnamefont {Sch{\"a}fer}},\ and\ \bibinfo {author} {\bibfnamefont
  {C.~A.}\ \bibnamefont {Stafford}},\ }\bibfield  {title} {\bibinfo {title}
  {Rabi oscillations at exceptional points in microwave billiards},\ }\href
  {https://doi.org/10.1103/PhysRevE.75.027201} {\bibfield  {journal} {\bibinfo
  {journal} {Phys. Rev. E}\ }\textbf {\bibinfo {volume} {75}},\ \bibinfo
  {pages} {027201} (\bibinfo {year} {2007})}\BibitemShut {NoStop}%
\bibitem [{\citenamefont {Lee}\ \emph {et~al.}(2009)\citenamefont {Lee},
  \citenamefont {Yang}, \citenamefont {Moon}, \citenamefont {Lee},
  \citenamefont {Shim}, \citenamefont {Kim}, \citenamefont {Lee},\ and\
  \citenamefont {An}}]{LYM09}%
  \BibitemOpen
  \bibfield  {author} {\bibinfo {author} {\bibfnamefont {S.-B.}\ \bibnamefont
  {Lee}}, \bibinfo {author} {\bibfnamefont {J.}~\bibnamefont {Yang}}, \bibinfo
  {author} {\bibfnamefont {S.}~\bibnamefont {Moon}}, \bibinfo {author}
  {\bibfnamefont {S.-Y.}\ \bibnamefont {Lee}}, \bibinfo {author} {\bibfnamefont
  {J.-B.}\ \bibnamefont {Shim}}, \bibinfo {author} {\bibfnamefont {S.~W.}\
  \bibnamefont {Kim}}, \bibinfo {author} {\bibfnamefont {J.-H.}\ \bibnamefont
  {Lee}},\ and\ \bibinfo {author} {\bibfnamefont {K.}~\bibnamefont {An}},\
  }\bibfield  {title} {\bibinfo {title} {Observation of an {E}xceptional
  {P}oint in a {C}haotic {O}ptical {M}icrocavity},\ }\href
  {https://doi.org/10.1103/PhysRevLett.103.134101} {\bibfield  {journal}
  {\bibinfo  {journal} {Phys. Rev. Lett.}\ }\textbf {\bibinfo {volume} {103}},\
  \bibinfo {pages} {134101} (\bibinfo {year} {2009})}\BibitemShut {NoStop}%
\bibitem [{\citenamefont {Peng}\ \emph
  {et~al.}(2014{\natexlab{a}})\citenamefont {Peng}, \citenamefont
  {{\"O}zdemir}, \citenamefont {Lei}, \citenamefont {Monfi}, \citenamefont
  {Gianfreda}, \citenamefont {Long}, \citenamefont {Fan}, \citenamefont {Nori},
  \citenamefont {Bender},\ and\ \citenamefont {Yang}}]{POL14}%
  \BibitemOpen
  \bibfield  {author} {\bibinfo {author} {\bibfnamefont {B.}~\bibnamefont
  {Peng}}, \bibinfo {author} {\bibfnamefont {{\c{S}}.~K.}\ \bibnamefont
  {{\"O}zdemir}}, \bibinfo {author} {\bibfnamefont {F.}~\bibnamefont {Lei}},
  \bibinfo {author} {\bibfnamefont {F.}~\bibnamefont {Monfi}}, \bibinfo
  {author} {\bibfnamefont {M.}~\bibnamefont {Gianfreda}}, \bibinfo {author}
  {\bibfnamefont {G.~L.}\ \bibnamefont {Long}}, \bibinfo {author}
  {\bibfnamefont {S.}~\bibnamefont {Fan}}, \bibinfo {author} {\bibfnamefont
  {F.}~\bibnamefont {Nori}}, \bibinfo {author} {\bibfnamefont {C.~M.}\
  \bibnamefont {Bender}},\ and\ \bibinfo {author} {\bibfnamefont
  {L.}~\bibnamefont {Yang}},\ }\bibfield  {title} {\bibinfo {title}
  {Parity-time-symmetric whispering-gallery microcavities},\ }\href
  {https://doi.org/10.1038/nphys2927} {\bibfield  {journal} {\bibinfo
  {journal} {Nature Physics}\ }\textbf {\bibinfo {volume} {10}},\ \bibinfo
  {pages} {394} (\bibinfo {year} {2014}{\natexlab{a}})}\BibitemShut {NoStop}%
\bibitem [{\citenamefont {Peng}\ \emph {et~al.}(2016)\citenamefont {Peng},
  \citenamefont {{\"O}zdemir}, \citenamefont {Liertzer}, \citenamefont {Chen},
  \citenamefont {Kramer}, \citenamefont {Yilmaz}, \citenamefont {Wiersig},
  \citenamefont {Rotter},\ and\ \citenamefont {Yang}}]{POL16}%
  \BibitemOpen
  \bibfield  {author} {\bibinfo {author} {\bibfnamefont {B.}~\bibnamefont
  {Peng}}, \bibinfo {author} {\bibfnamefont {{\c{S}}.~K.}\ \bibnamefont
  {{\"O}zdemir}}, \bibinfo {author} {\bibfnamefont {M.}~\bibnamefont
  {Liertzer}}, \bibinfo {author} {\bibfnamefont {W.}~\bibnamefont {Chen}},
  \bibinfo {author} {\bibfnamefont {J.}~\bibnamefont {Kramer}}, \bibinfo
  {author} {\bibfnamefont {H.}~\bibnamefont {Yilmaz}}, \bibinfo {author}
  {\bibfnamefont {J.}~\bibnamefont {Wiersig}}, \bibinfo {author} {\bibfnamefont
  {S.}~\bibnamefont {Rotter}},\ and\ \bibinfo {author} {\bibfnamefont
  {L.}~\bibnamefont {Yang}},\ }\bibfield  {title} {\bibinfo {title} {Chiral
  modes and directional lasing at exceptional points},\ }\href
  {https://doi.org/10.1073/pnas.1603318113} {\bibfield  {journal} {\bibinfo
  {journal} {Proc. Natl. Acad. Sci. USA}\ }\textbf {\bibinfo {volume} {113}},\
  \bibinfo {pages} {6845} (\bibinfo {year} {2016})}\BibitemShut {NoStop}%
\bibitem [{\citenamefont {Richter}\ \emph {et~al.}(2019)\citenamefont
  {Richter}, \citenamefont {Zirnstein}, \citenamefont
  {Z\'{u}\~{n}iga{-}P\'{e}rez}, \citenamefont {Kr{\"u}ger}, \citenamefont
  {Deparis}, \citenamefont {Trefflich}, \citenamefont {Sturm}, \citenamefont
  {Rosenow}, \citenamefont {Grundmann},\ and\ \citenamefont
  {Schmidt-Grund}}]{RZZ19}%
  \BibitemOpen
  \bibfield  {author} {\bibinfo {author} {\bibfnamefont {S.}~\bibnamefont
  {Richter}}, \bibinfo {author} {\bibfnamefont {H.-G.}\ \bibnamefont
  {Zirnstein}}, \bibinfo {author} {\bibfnamefont {J.}~\bibnamefont
  {Z\'{u}\~{n}iga{-}P\'{e}rez}}, \bibinfo {author} {\bibfnamefont
  {E.}~\bibnamefont {Kr{\"u}ger}}, \bibinfo {author} {\bibfnamefont
  {C.}~\bibnamefont {Deparis}}, \bibinfo {author} {\bibfnamefont
  {L.}~\bibnamefont {Trefflich}}, \bibinfo {author} {\bibfnamefont
  {C.}~\bibnamefont {Sturm}}, \bibinfo {author} {\bibfnamefont
  {B.}~\bibnamefont {Rosenow}}, \bibinfo {author} {\bibfnamefont
  {M.}~\bibnamefont {Grundmann}},\ and\ \bibinfo {author} {\bibfnamefont
  {R.}~\bibnamefont {Schmidt-Grund}},\ }\bibfield  {title} {\bibinfo {title}
  {Voigt {E}xceptional {P}oints in an {A}nisotropic {Z}n{O}-{B}ased {P}lanar
  {M}icrocavity: {S}quare-{R}oot {T}opology, {P}olarization {V}ortices, and
  {C}ircularity},\ }\href {https://doi.org/10.1103/PhysRevLett.123.227401}
  {\bibfield  {journal} {\bibinfo  {journal} {Phys. Rev. Lett.}\ }\textbf
  {\bibinfo {volume} {123}},\ \bibinfo {pages} {227401} (\bibinfo {year}
  {2019})}\BibitemShut {NoStop}%
\bibitem [{\citenamefont {Choi}\ \emph {et~al.}(2010)\citenamefont {Choi},
  \citenamefont {Kang}, \citenamefont {Lim}, \citenamefont {Kim}, \citenamefont
  {Kim}, \citenamefont {Lee},\ and\ \citenamefont {An}}]{CKL10}%
  \BibitemOpen
  \bibfield  {author} {\bibinfo {author} {\bibfnamefont {Y.}~\bibnamefont
  {Choi}}, \bibinfo {author} {\bibfnamefont {S.}~\bibnamefont {Kang}}, \bibinfo
  {author} {\bibfnamefont {S.}~\bibnamefont {Lim}}, \bibinfo {author}
  {\bibfnamefont {W.}~\bibnamefont {Kim}}, \bibinfo {author} {\bibfnamefont
  {J.-R.}\ \bibnamefont {Kim}}, \bibinfo {author} {\bibfnamefont {J.-H.}\
  \bibnamefont {Lee}},\ and\ \bibinfo {author} {\bibfnamefont {K.}~\bibnamefont
  {An}},\ }\bibfield  {title} {\bibinfo {title} {Quasieigenstate {C}oalescence
  in an {A}tom-{C}avity {Q}uantum {C}omposite},\ }\href
  {https://doi.org/10.1103/PhysRevLett.104.153601} {\bibfield  {journal}
  {\bibinfo  {journal} {Phys. Rev. Lett.}\ }\textbf {\bibinfo {volume} {104}},\
  \bibinfo {pages} {153601} (\bibinfo {year} {2010})}\BibitemShut {NoStop}%
\bibitem [{\citenamefont {Regensburger}\ \emph {et~al.}(2012)\citenamefont
  {Regensburger}, \citenamefont {Bersch}, \citenamefont {Miri}, \citenamefont
  {Onishchukov}, \citenamefont {Christodoulides},\ and\ \citenamefont
  {Peschel}}]{RBM12}%
  \BibitemOpen
  \bibfield  {author} {\bibinfo {author} {\bibfnamefont {A.}~\bibnamefont
  {Regensburger}}, \bibinfo {author} {\bibfnamefont {C.}~\bibnamefont
  {Bersch}}, \bibinfo {author} {\bibfnamefont {M.-A.}\ \bibnamefont {Miri}},
  \bibinfo {author} {\bibfnamefont {G.}~\bibnamefont {Onishchukov}}, \bibinfo
  {author} {\bibfnamefont {D.~N.}\ \bibnamefont {Christodoulides}},\ and\
  \bibinfo {author} {\bibfnamefont {U.}~\bibnamefont {Peschel}},\ }\bibfield
  {title} {\bibinfo {title} {Parity-time synthetic photonic lattices},\ }\href
  {https://doi.org/10.1038/nature11298} {\bibfield  {journal} {\bibinfo
  {journal} {Nature (London)}\ }\textbf {\bibinfo {volume} {488}},\ \bibinfo
  {pages} {167–171} (\bibinfo {year} {2012})}\BibitemShut {NoStop}%
\bibitem [{\citenamefont {Gao}\ \emph {et~al.}(2015)\citenamefont {Gao},
  \citenamefont {Estrecho}, \citenamefont {Bliokh}, \citenamefont {Liev},
  \citenamefont {Fraser}, \citenamefont {Brodbeck}, \citenamefont {Kamp},
  \citenamefont {Schneider}, \citenamefont {H{\"o}fling}, \citenamefont
  {Yamamoto}, \citenamefont {Nori}, \citenamefont {Kivshar}, \citenamefont
  {Truscott}, \citenamefont {Dall},\ and\ \citenamefont {Ostrovskaya}}]{GEB15}%
  \BibitemOpen
  \bibfield  {author} {\bibinfo {author} {\bibfnamefont {T.}~\bibnamefont
  {Gao}}, \bibinfo {author} {\bibfnamefont {E.}~\bibnamefont {Estrecho}},
  \bibinfo {author} {\bibfnamefont {K.~Y.}\ \bibnamefont {Bliokh}}, \bibinfo
  {author} {\bibfnamefont {T.~C.~H.}\ \bibnamefont {Liev}}, \bibinfo {author}
  {\bibfnamefont {M.~D.}\ \bibnamefont {Fraser}}, \bibinfo {author}
  {\bibfnamefont {S.}~\bibnamefont {Brodbeck}}, \bibinfo {author}
  {\bibfnamefont {M.}~\bibnamefont {Kamp}}, \bibinfo {author} {\bibfnamefont
  {C.}~\bibnamefont {Schneider}}, \bibinfo {author} {\bibfnamefont
  {S.}~\bibnamefont {H{\"o}fling}}, \bibinfo {author} {\bibfnamefont
  {Y.}~\bibnamefont {Yamamoto}}, \bibinfo {author} {\bibfnamefont
  {F.}~\bibnamefont {Nori}}, \bibinfo {author} {\bibfnamefont {Y.~S.}\
  \bibnamefont {Kivshar}}, \bibinfo {author} {\bibfnamefont {A.~G.}\
  \bibnamefont {Truscott}}, \bibinfo {author} {\bibfnamefont {R.~G.}\
  \bibnamefont {Dall}},\ and\ \bibinfo {author} {\bibfnamefont {E.~A.}\
  \bibnamefont {Ostrovskaya}},\ }\bibfield  {title} {\bibinfo {title}
  {Observation of non-{H}ermitian degeneracies in a chaotic exciton-polariton
  billiard},\ }\href {https://doi.org/10.1038/nature15522} {\bibfield
  {journal} {\bibinfo  {journal} {Nature (London)}\ }\textbf {\bibinfo {volume}
  {526}},\ \bibinfo {pages} {554} (\bibinfo {year} {2015})}\BibitemShut
  {NoStop}%
\bibitem [{\citenamefont {Shin}\ \emph {et~al.}(2016)\citenamefont {Shin},
  \citenamefont {Kwak}, \citenamefont {Moon}, \citenamefont {Lee},
  \citenamefont {Yang},\ and\ \citenamefont {An}}]{SKM16}%
  \BibitemOpen
  \bibfield  {author} {\bibinfo {author} {\bibfnamefont {Y.}~\bibnamefont
  {Shin}}, \bibinfo {author} {\bibfnamefont {H.}~\bibnamefont {Kwak}}, \bibinfo
  {author} {\bibfnamefont {S.}~\bibnamefont {Moon}}, \bibinfo {author}
  {\bibfnamefont {S.-B.}\ \bibnamefont {Lee}}, \bibinfo {author} {\bibfnamefont
  {J.}~\bibnamefont {Yang}},\ and\ \bibinfo {author} {\bibfnamefont
  {K.}~\bibnamefont {An}},\ }\bibfield  {title} {\bibinfo {title} {Observation
  of an exceptional point in a two-dimensional ultrasonic cavity of concentric
  circular shells},\ }\href {https://doi.org/10.1038/srep38826} {\bibfield
  {journal} {\bibinfo  {journal} {Sci. Rep.}\ }\textbf {\bibinfo {volume}
  {6}},\ \bibinfo {pages} {38826} (\bibinfo {year} {2016})}\BibitemShut
  {NoStop}%
\bibitem [{\citenamefont {Wang}\ \emph {et~al.}(2019)\citenamefont {Wang},
  \citenamefont {Hou}, \citenamefont {Lu}, \citenamefont {Chen}, \citenamefont
  {Zhang},\ and\ \citenamefont {Chan}}]{WHL19}%
  \BibitemOpen
  \bibfield  {author} {\bibinfo {author} {\bibfnamefont {S.}~\bibnamefont
  {Wang}}, \bibinfo {author} {\bibfnamefont {B.}~\bibnamefont {Hou}}, \bibinfo
  {author} {\bibfnamefont {W.}~\bibnamefont {Lu}}, \bibinfo {author}
  {\bibfnamefont {Y.}~\bibnamefont {Chen}}, \bibinfo {author} {\bibfnamefont
  {Z.~Q.}\ \bibnamefont {Zhang}},\ and\ \bibinfo {author} {\bibfnamefont
  {C.~T.}\ \bibnamefont {Chan}},\ }\bibfield  {title} {\bibinfo {title}
  {Arbitrary order exceptional point induced by photonic spin–orbit
  interaction in coupled resonators},\ }\href
  {https://doi.org/10.1038/s41467-019-08826-6} {\bibfield  {journal} {\bibinfo
  {journal} {Nat. Commun.}\ }\textbf {\bibinfo {volume} {10}},\ \bibinfo
  {pages} {832} (\bibinfo {year} {2019})}\BibitemShut {NoStop}%
\bibitem [{\citenamefont {Miao}\ \emph {et~al.}(2016)\citenamefont {Miao},
  \citenamefont {Zhang}, \citenamefont {Sun}, \citenamefont {Walasik},
  \citenamefont {Longhi}, \citenamefont {Litchinitser},\ and\ \citenamefont
  {Feng}}]{MZS16}%
  \BibitemOpen
  \bibfield  {author} {\bibinfo {author} {\bibfnamefont {P.}~\bibnamefont
  {Miao}}, \bibinfo {author} {\bibfnamefont {Z.}~\bibnamefont {Zhang}},
  \bibinfo {author} {\bibfnamefont {J.}~\bibnamefont {Sun}}, \bibinfo {author}
  {\bibfnamefont {W.}~\bibnamefont {Walasik}}, \bibinfo {author} {\bibfnamefont
  {S.}~\bibnamefont {Longhi}}, \bibinfo {author} {\bibfnamefont {N.~M.}\
  \bibnamefont {Litchinitser}},\ and\ \bibinfo {author} {\bibfnamefont
  {L.}~\bibnamefont {Feng}},\ }\bibfield  {title} {\bibinfo {title} {Orbital
  angular momentum microlaser},\ }\href
  {https://doi.org/10.1126/science.aaf8533} {\bibfield  {journal} {\bibinfo
  {journal} {Science}\ }\textbf {\bibinfo {volume} {353}},\ \bibinfo {pages}
  {464} (\bibinfo {year} {2016})}\BibitemShut {NoStop}%
\bibitem [{\citenamefont {Richter}\ \emph {et~al.}(2017)\citenamefont
  {Richter}, \citenamefont {Michalsky}, \citenamefont {Sturm}, \citenamefont
  {Rosenow}, \citenamefont {Grundmann},\ and\ \citenamefont
  {Schmidt-Grund}}]{RMS17}%
  \BibitemOpen
  \bibfield  {author} {\bibinfo {author} {\bibfnamefont {S.}~\bibnamefont
  {Richter}}, \bibinfo {author} {\bibfnamefont {T.}~\bibnamefont {Michalsky}},
  \bibinfo {author} {\bibfnamefont {C.}~\bibnamefont {Sturm}}, \bibinfo
  {author} {\bibfnamefont {B.}~\bibnamefont {Rosenow}}, \bibinfo {author}
  {\bibfnamefont {M.}~\bibnamefont {Grundmann}},\ and\ \bibinfo {author}
  {\bibfnamefont {R.}~\bibnamefont {Schmidt-Grund}},\ }\bibfield  {title}
  {\bibinfo {title} {Exceptional points in anisotropic planar microcavities},\
  }\href {https://doi.org/10.1103/PhysRevA.95.023836} {\bibfield  {journal}
  {\bibinfo  {journal} {Phys. Rev. A}\ }\textbf {\bibinfo {volume} {95}},\
  \bibinfo {pages} {023836} (\bibinfo {year} {2017})}\BibitemShut {NoStop}%
\bibitem [{\citenamefont {Xu}\ \emph {et~al.}(2016)\citenamefont {Xu},
  \citenamefont {Mason}, \citenamefont {Jiang},\ and\ \citenamefont
  {Harris}}]{XMJ16}%
  \BibitemOpen
  \bibfield  {author} {\bibinfo {author} {\bibfnamefont {H.}~\bibnamefont
  {Xu}}, \bibinfo {author} {\bibfnamefont {D.}~\bibnamefont {Mason}}, \bibinfo
  {author} {\bibfnamefont {L.}~\bibnamefont {Jiang}},\ and\ \bibinfo {author}
  {\bibfnamefont {J.~G.~E.}\ \bibnamefont {Harris}},\ }\bibfield  {title}
  {\bibinfo {title} {Topological energy transfer in an optomechanical system
  with exceptional points},\ }\href {https://doi.org/10.1038/nature18604}
  {\bibfield  {journal} {\bibinfo  {journal} {Nature (London)}\ }\textbf
  {\bibinfo {volume} {537}},\ \bibinfo {pages} {80} (\bibinfo {year}
  {2016})}\BibitemShut {NoStop}%
\bibitem [{\citenamefont {Doppler}\ \emph {et~al.}(2016)\citenamefont
  {Doppler}, \citenamefont {Mailybaev}, \citenamefont {B{\"o}hm}, \citenamefont
  {Kuhl}, \citenamefont {Girschik}, \citenamefont {Libisch}, \citenamefont
  {Milburn}, \citenamefont {Rabl}, \citenamefont {Moiseyev},\ and\
  \citenamefont {Rotter}}]{DMB16}%
  \BibitemOpen
  \bibfield  {author} {\bibinfo {author} {\bibfnamefont {J.}~\bibnamefont
  {Doppler}}, \bibinfo {author} {\bibfnamefont {A.~A.}\ \bibnamefont
  {Mailybaev}}, \bibinfo {author} {\bibfnamefont {J.}~\bibnamefont {B{\"o}hm}},
  \bibinfo {author} {\bibfnamefont {U.}~\bibnamefont {Kuhl}}, \bibinfo {author}
  {\bibfnamefont {A.}~\bibnamefont {Girschik}}, \bibinfo {author}
  {\bibfnamefont {F.}~\bibnamefont {Libisch}}, \bibinfo {author} {\bibfnamefont
  {T.~J.}\ \bibnamefont {Milburn}}, \bibinfo {author} {\bibfnamefont
  {P.}~\bibnamefont {Rabl}}, \bibinfo {author} {\bibfnamefont {N.}~\bibnamefont
  {Moiseyev}},\ and\ \bibinfo {author} {\bibfnamefont {S.}~\bibnamefont
  {Rotter}},\ }\bibfield  {title} {\bibinfo {title} {Dynamically encircling an
  exceptional point for asymmetric mode switching},\ }\href
  {https://doi.org/10.1038/nature18605} {\bibfield  {journal} {\bibinfo
  {journal} {Nature (London)}\ }\textbf {\bibinfo {volume} {537}},\ \bibinfo
  {pages} {76} (\bibinfo {year} {2016})}\BibitemShut {NoStop}%
\bibitem [{\citenamefont {Peng}\ \emph
  {et~al.}(2014{\natexlab{b}})\citenamefont {Peng}, \citenamefont
  {{\"O}zdemir}, \citenamefont {Rotter}, \citenamefont {Y{\i}lmaz},
  \citenamefont {Liertzer}, \citenamefont {Monfi}, \citenamefont {Bender},
  \citenamefont {Nori},\ and\ \citenamefont {Yang}}]{POR14}%
  \BibitemOpen
  \bibfield  {author} {\bibinfo {author} {\bibfnamefont {B.}~\bibnamefont
  {Peng}}, \bibinfo {author} {\bibfnamefont {{\c{S}}.~K.}\ \bibnamefont
  {{\"O}zdemir}}, \bibinfo {author} {\bibfnamefont {S.}~\bibnamefont {Rotter}},
  \bibinfo {author} {\bibfnamefont {H.}~\bibnamefont {Y{\i}lmaz}}, \bibinfo
  {author} {\bibfnamefont {M.}~\bibnamefont {Liertzer}}, \bibinfo {author}
  {\bibfnamefont {F.}~\bibnamefont {Monfi}}, \bibinfo {author} {\bibfnamefont
  {C.~M.}\ \bibnamefont {Bender}}, \bibinfo {author} {\bibfnamefont
  {F.}~\bibnamefont {Nori}},\ and\ \bibinfo {author} {\bibfnamefont
  {L.}~\bibnamefont {Yang}},\ }\bibfield  {title} {\bibinfo {title}
  {Loss-induced suppression and revival of lasing},\ }\href
  {https://doi.org/10.1126/science.1258004} {\bibfield  {journal} {\bibinfo
  {journal} {Science}\ }\textbf {\bibinfo {volume} {17}},\ \bibinfo {pages}
  {328} (\bibinfo {year} {2014}{\natexlab{b}})}\BibitemShut {NoStop}%
\bibitem [{\citenamefont {Hodaei}\ \emph {et~al.}(2014)\citenamefont {Hodaei},
  \citenamefont {Miri}, \citenamefont {Heinrich}, \citenamefont
  {Christodoulides},\ and\ \citenamefont {Khajavikhan}}]{HMH14}%
  \BibitemOpen
  \bibfield  {author} {\bibinfo {author} {\bibfnamefont {H.}~\bibnamefont
  {Hodaei}}, \bibinfo {author} {\bibfnamefont {M.-A.}\ \bibnamefont {Miri}},
  \bibinfo {author} {\bibfnamefont {M.}~\bibnamefont {Heinrich}}, \bibinfo
  {author} {\bibfnamefont {D.}~\bibnamefont {Christodoulides}},\ and\ \bibinfo
  {author} {\bibfnamefont {M.}~\bibnamefont {Khajavikhan}},\ }\bibfield
  {title} {\bibinfo {title} {Parity-time-symmetric microring lasers},\ }\href
  {https://doi.org/10.1126/science.1258480} {\bibfield  {journal} {\bibinfo
  {journal} {Science}\ }\textbf {\bibinfo {volume} {346}},\ \bibinfo {pages}
  {975} (\bibinfo {year} {2014})}\BibitemShut {NoStop}%
\bibitem [{\citenamefont {Wiersig}(2014)}]{Wiersig14b}%
  \BibitemOpen
  \bibfield  {author} {\bibinfo {author} {\bibfnamefont {J.}~\bibnamefont
  {Wiersig}},\ }\bibfield  {title} {\bibinfo {title} {Enhancing the
  {S}ensitivity of {F}requency and {E}nergy {S}plitting {D}etection by {U}sing
  {E}xceptional {P}oints: {A}pplication to {M}icrocavity {S}ensors for
  {S}ingle-{P}article {D}etection},\ }\href
  {https://doi.org/10.1103/PhysRevLett.112.203901} {\bibfield  {journal}
  {\bibinfo  {journal} {Phys. Rev. Lett.}\ }\textbf {\bibinfo {volume} {112}},\
  \bibinfo {pages} {203901} (\bibinfo {year} {2014})}\BibitemShut {NoStop}%
\bibitem [{\citenamefont {Chen}\ \emph {et~al.}(2017)\citenamefont {Chen},
  \citenamefont {{\"O}zdemir}, \citenamefont {Zhao}, \citenamefont {Wiersig},\
  and\ \citenamefont {Yang}}]{COZ17}%
  \BibitemOpen
  \bibfield  {author} {\bibinfo {author} {\bibfnamefont {W.}~\bibnamefont
  {Chen}}, \bibinfo {author} {\bibfnamefont {{\c{S}}.~K.}\ \bibnamefont
  {{\"O}zdemir}}, \bibinfo {author} {\bibfnamefont {G.}~\bibnamefont {Zhao}},
  \bibinfo {author} {\bibfnamefont {J.}~\bibnamefont {Wiersig}},\ and\ \bibinfo
  {author} {\bibfnamefont {L.}~\bibnamefont {Yang}},\ }\bibfield  {title}
  {\bibinfo {title} {Exceptional points enhance sensing in an optical
  microcavity},\ }\href {https://doi.org/10.1038/nature23281} {\bibfield
  {journal} {\bibinfo  {journal} {Nature (London)}\ }\textbf {\bibinfo {volume}
  {548}},\ \bibinfo {pages} {192} (\bibinfo {year} {2017})}\BibitemShut
  {NoStop}%
\bibitem [{\citenamefont {Hodaei}\ \emph {et~al.}(2017)\citenamefont {Hodaei},
  \citenamefont {Hassan}, \citenamefont {Wittek}, \citenamefont
  {Carcia-Cracia}, \citenamefont {El-Ganainy}, \citenamefont
  {Christodoulides},\ and\ \citenamefont {Khajavikhan}}]{HHW17}%
  \BibitemOpen
  \bibfield  {author} {\bibinfo {author} {\bibfnamefont {H.}~\bibnamefont
  {Hodaei}}, \bibinfo {author} {\bibfnamefont {A.}~\bibnamefont {Hassan}},
  \bibinfo {author} {\bibfnamefont {S.}~\bibnamefont {Wittek}}, \bibinfo
  {author} {\bibfnamefont {H.}~\bibnamefont {Carcia-Cracia}}, \bibinfo {author}
  {\bibfnamefont {R.}~\bibnamefont {El-Ganainy}}, \bibinfo {author}
  {\bibfnamefont {D.}~\bibnamefont {Christodoulides}},\ and\ \bibinfo {author}
  {\bibfnamefont {M.}~\bibnamefont {Khajavikhan}},\ }\bibfield  {title}
  {\bibinfo {title} {Enhanced sensitivity at higher-order exceptional points},\
  }\href {https://doi.org/10.1038/nature23280} {\bibfield  {journal} {\bibinfo
  {journal} {Nature (London)}\ }\textbf {\bibinfo {volume} {548}},\ \bibinfo
  {pages} {187} (\bibinfo {year} {2017})}\BibitemShut {NoStop}%
\bibitem [{\citenamefont {Xiao}\ \emph {et~al.}(2019)\citenamefont {Xiao},
  \citenamefont {Li}, \citenamefont {Kottos},\ and\ \citenamefont
  {Al{\`u}}}]{XLK19}%
  \BibitemOpen
  \bibfield  {author} {\bibinfo {author} {\bibfnamefont {Z.}~\bibnamefont
  {Xiao}}, \bibinfo {author} {\bibfnamefont {H.}~\bibnamefont {Li}}, \bibinfo
  {author} {\bibfnamefont {T.}~\bibnamefont {Kottos}},\ and\ \bibinfo {author}
  {\bibfnamefont {A.}~\bibnamefont {Al{\`u}}},\ }\bibfield  {title} {\bibinfo
  {title} {Enhanced sensing and nondegraded thermal noise performance based on
  {PT}-symmetric electronic circuits with a sixth-order exceptional point},\
  }\href {https://doi.org/10.1103/PhysRevLett.123.213901} {\bibfield  {journal}
  {\bibinfo  {journal} {Phys. Rev. Lett.}\ }\textbf {\bibinfo {volume} {123}},\
  \bibinfo {pages} {213901} (\bibinfo {year} {2019})}\BibitemShut {NoStop}%
\bibitem [{\citenamefont {Lai}\ \emph {et~al.}(2019)\citenamefont {Lai},
  \citenamefont {Lu}, \citenamefont {Suh}, \citenamefont {Yuan},\ and\
  \citenamefont {Vahala}}]{LLS19}%
  \BibitemOpen
  \bibfield  {author} {\bibinfo {author} {\bibfnamefont {Y.-H.}\ \bibnamefont
  {Lai}}, \bibinfo {author} {\bibfnamefont {Y.-K.}\ \bibnamefont {Lu}},
  \bibinfo {author} {\bibfnamefont {M.-G.}\ \bibnamefont {Suh}}, \bibinfo
  {author} {\bibfnamefont {Z.}~\bibnamefont {Yuan}},\ and\ \bibinfo {author}
  {\bibfnamefont {K.}~\bibnamefont {Vahala}},\ }\bibfield  {title} {\bibinfo
  {title} {Observation of the exceptional-point-enhanced {S}agnac effect},\
  }\href {https://doi.org/10.1038/s41586-019-1777-z} {\bibfield  {journal}
  {\bibinfo  {journal} {Nature (London)}\ }\textbf {\bibinfo {volume} {576}},\
  \bibinfo {pages} {65} (\bibinfo {year} {2019})}\BibitemShut {NoStop}%
\bibitem [{\citenamefont {Wang}\ \emph {et~al.}(2020)\citenamefont {Wang},
  \citenamefont {Lai}, \citenamefont {Yuan}, \citenamefont {Suh},\ and\
  \citenamefont {Vahala}}]{WLY20}%
  \BibitemOpen
  \bibfield  {author} {\bibinfo {author} {\bibfnamefont {H.}~\bibnamefont
  {Wang}}, \bibinfo {author} {\bibfnamefont {Y.-H.}\ \bibnamefont {Lai}},
  \bibinfo {author} {\bibfnamefont {Z.}~\bibnamefont {Yuan}}, \bibinfo {author}
  {\bibfnamefont {M.-G.}\ \bibnamefont {Suh}},\ and\ \bibinfo {author}
  {\bibfnamefont {K.}~\bibnamefont {Vahala}},\ }\bibfield  {title} {\bibinfo
  {title} {Petermann-factor sensitivity limit near an exceptional point in a
  {B}rillouin ring laser gyroscope},\ }\href
  {https://doi.org/10.1038/s41467-020-15341-6} {\bibfield  {journal} {\bibinfo
  {journal} {Nat. Commun.}\ }\textbf {\bibinfo {volume} {11}},\ \bibinfo
  {pages} {1610} (\bibinfo {year} {2020})}\BibitemShut {NoStop}%
\bibitem [{\citenamefont {Kononchuk}\ \emph {et~al.}(2022)\citenamefont
  {Kononchuk}, \citenamefont {Cai}, \citenamefont {Ellis}, \citenamefont
  {Thevamaran},\ and\ \citenamefont {Kottos}}]{KCE22}%
  \BibitemOpen
  \bibfield  {author} {\bibinfo {author} {\bibfnamefont {R.}~\bibnamefont
  {Kononchuk}}, \bibinfo {author} {\bibfnamefont {J.}~\bibnamefont {Cai}},
  \bibinfo {author} {\bibfnamefont {F.}~\bibnamefont {Ellis}}, \bibinfo
  {author} {\bibfnamefont {R.}~\bibnamefont {Thevamaran}},\ and\ \bibinfo
  {author} {\bibfnamefont {T.}~\bibnamefont {Kottos}},\ }\bibfield  {title}
  {\bibinfo {title} {Exceptional-point-based accelerometers with enhanced
  signal-to-noise ratio},\ }\href {https://doi.org/10.1038/s41586-022-04904-w}
  {\bibfield  {journal} {\bibinfo  {journal} {Nature (London)}\ }\textbf
  {\bibinfo {volume} {607}},\ \bibinfo {pages} {697} (\bibinfo {year}
  {2022})}\BibitemShut {NoStop}%
\bibitem [{\citenamefont {Wiersig}(2020)}]{Wiersig20c}%
  \BibitemOpen
  \bibfield  {author} {\bibinfo {author} {\bibfnamefont {J.}~\bibnamefont
  {Wiersig}},\ }\bibfield  {title} {\bibinfo {title} {Review of exceptional
  point-based sensors},\ }\href {https://doi.org/10.1364/PRJ.396115} {\bibfield
   {journal} {\bibinfo  {journal} {Photonics Res.}\ }\textbf {\bibinfo {volume}
  {8}},\ \bibinfo {pages} {1457} (\bibinfo {year} {2020})}\BibitemShut
  {NoStop}%
\bibitem [{\citenamefont {Wiersig}(2022)}]{Wiersig22}%
  \BibitemOpen
  \bibfield  {author} {\bibinfo {author} {\bibfnamefont {J.}~\bibnamefont
  {Wiersig}},\ }\bibfield  {title} {\bibinfo {title} {Response strengths of
  open systems at exceptional points},\ }\href
  {https://doi.org/10.1103/PhysRevResearch.4.023121} {\bibfield  {journal}
  {\bibinfo  {journal} {Phys. Rev. Res.}\ }\textbf {\bibinfo {volume} {4}},\
  \bibinfo {pages} {023121} (\bibinfo {year} {2022})}\BibitemShut {NoStop}%
\bibitem [{\citenamefont {Keck}\ \emph {et~al.}(2003)\citenamefont {Keck},
  \citenamefont {Korsch},\ and\ \citenamefont {Mossmann}}]{KKM03}%
  \BibitemOpen
  \bibfield  {author} {\bibinfo {author} {\bibfnamefont {F.}~\bibnamefont
  {Keck}}, \bibinfo {author} {\bibfnamefont {H.~J.}\ \bibnamefont {Korsch}},\
  and\ \bibinfo {author} {\bibfnamefont {S.}~\bibnamefont {Mossmann}},\
  }\bibfield  {title} {\bibinfo {title} {Unfolding a diabolic point: a
  generalized crossing scenario},\ }\href
  {https://iopscience.iop.org/article/10.1088/0305-4470/36/8/310} {\bibfield
  {journal} {\bibinfo  {journal} {J. Phys. A: Math. Gen.}\ }\textbf {\bibinfo
  {volume} {36}},\ \bibinfo {pages} {2125} (\bibinfo {year}
  {2003})}\BibitemShut {NoStop}%
\bibitem [{\citenamefont {Wang}\ \emph {et~al.}(2021)\citenamefont {Wang},
  \citenamefont {Zhang},\ and\ \citenamefont {Song}}]{WZS21}%
  \BibitemOpen
  \bibfield  {author} {\bibinfo {author} {\bibfnamefont {P.}~\bibnamefont
  {Wang}}, \bibinfo {author} {\bibfnamefont {K.~L.}\ \bibnamefont {Zhang}},\
  and\ \bibinfo {author} {\bibfnamefont {Z.}~\bibnamefont {Song}},\ }\bibfield
  {title} {\bibinfo {title} {Transition from degeneracy to coalescence:
  {T}heorem and applications},\ }\href
  {https://doi.org/10.1103/PhysRevB.104.245406} {\bibfield  {journal} {\bibinfo
   {journal} {Phys. Rev. B}\ }\textbf {\bibinfo {volume} {104}},\ \bibinfo
  {pages} {245406} (\bibinfo {year} {2021})}\BibitemShut {NoStop}%
\bibitem [{\citenamefont {Wiersig}(2011)}]{Wiersig11}%
  \BibitemOpen
  \bibfield  {author} {\bibinfo {author} {\bibfnamefont {J.}~\bibnamefont
  {Wiersig}},\ }\bibfield  {title} {\bibinfo {title} {Structure of
  whispering-gallery modes in optical microdisks perturbed by nanoparticles},\
  }\href {https://doi.org/10.1103/PhysRevA.84.063828} {\bibfield  {journal}
  {\bibinfo  {journal} {Phys. Rev. A}\ }\textbf {\bibinfo {volume} {84}},\
  \bibinfo {pages} {063828} (\bibinfo {year} {2011})}\BibitemShut {NoStop}%
\bibitem [{\citenamefont {Wiersig}(2018)}]{Wiersig18b}%
  \BibitemOpen
  \bibfield  {author} {\bibinfo {author} {\bibfnamefont {J.}~\bibnamefont
  {Wiersig}},\ }\bibfield  {title} {\bibinfo {title} {Non-{H}ermitian effects
  due to asymmetric backscattering of light in whispering-gallery
  microcavities},\ }in\ \href
  {https://link.springer.com/chapter/10.1007/978-981-13-1247-2_6} {\emph
  {\bibinfo {booktitle} {Parity-time Symmetry and Its Applications}}},\
  \bibinfo {editor} {edited by\ \bibinfo {editor} {\bibfnamefont
  {D.}~\bibnamefont {Christodoulides}}\ and\ \bibinfo {editor} {\bibfnamefont
  {J.}~\bibnamefont {Yang}}}\ (\bibinfo  {publisher} {Springer},\ \bibinfo
  {address} {Singapore},\ \bibinfo {year} {2018})\ pp.\ \bibinfo {pages}
  {155--184}\BibitemShut {NoStop}%
\bibitem [{\citenamefont {Yi}\ \emph {et~al.}(2018)\citenamefont {Yi},
  \citenamefont {Kullig},\ and\ \citenamefont {Wiersig}}]{YKW18}%
  \BibitemOpen
  \bibfield  {author} {\bibinfo {author} {\bibfnamefont {C.-H.}\ \bibnamefont
  {Yi}}, \bibinfo {author} {\bibfnamefont {J.}~\bibnamefont {Kullig}},\ and\
  \bibinfo {author} {\bibfnamefont {J.}~\bibnamefont {Wiersig}},\ }\bibfield
  {title} {\bibinfo {title} {Pair of exceptional points in a microdisk cavity
  under an extremely weak deformation},\ }\href
  {https://doi.org/10.1103/PhysRevLett.120.093902} {\bibfield  {journal}
  {\bibinfo  {journal} {Phys. Rev. Lett.}\ }\textbf {\bibinfo {volume} {120}},\
  \bibinfo {pages} {093902} (\bibinfo {year} {2018})}\BibitemShut {NoStop}%
\bibitem [{\citenamefont {Kullig}\ \emph {et~al.}(2018)\citenamefont {Kullig},
  \citenamefont {Yi},\ and\ \citenamefont {Wiersig}}]{KYW18}%
  \BibitemOpen
  \bibfield  {author} {\bibinfo {author} {\bibfnamefont {J.}~\bibnamefont
  {Kullig}}, \bibinfo {author} {\bibfnamefont {C.-H.}\ \bibnamefont {Yi}},\
  and\ \bibinfo {author} {\bibfnamefont {J.}~\bibnamefont {Wiersig}},\
  }\bibfield  {title} {\bibinfo {title} {Exceptional points by coupling of
  modes with different angular momenta in deformed microdisks: A perturbative
  analysis},\ }\href {https://doi.org/10.1103/PhysRevA.98.023851} {\bibfield
  {journal} {\bibinfo  {journal} {Phys. Rev. A}\ }\textbf {\bibinfo {volume}
  {98}},\ \bibinfo {pages} {023851} (\bibinfo {year} {2018})}\BibitemShut
  {NoStop}%
\bibitem [{\citenamefont {Feilhauer}\ \emph {et~al.}(2020)\citenamefont
  {Feilhauer}, \citenamefont {Schumer}, \citenamefont {Doppler}, \citenamefont
  {Mailybaev}, \citenamefont {B{\"o}hm}, \citenamefont {Kuhl}, \citenamefont
  {Moiseyev},\ and\ \citenamefont {Rotter}}]{FSD20}%
  \BibitemOpen
  \bibfield  {author} {\bibinfo {author} {\bibfnamefont {J.}~\bibnamefont
  {Feilhauer}}, \bibinfo {author} {\bibfnamefont {A.}~\bibnamefont {Schumer}},
  \bibinfo {author} {\bibfnamefont {J.}~\bibnamefont {Doppler}}, \bibinfo
  {author} {\bibfnamefont {A.~A.}\ \bibnamefont {Mailybaev}}, \bibinfo {author}
  {\bibfnamefont {J.}~\bibnamefont {B{\"o}hm}}, \bibinfo {author}
  {\bibfnamefont {U.}~\bibnamefont {Kuhl}}, \bibinfo {author} {\bibfnamefont
  {N.}~\bibnamefont {Moiseyev}},\ and\ \bibinfo {author} {\bibfnamefont
  {S.}~\bibnamefont {Rotter}},\ }\bibfield  {title} {\bibinfo {title}
  {Encircling exceptional points as a non-{H}ermitian extension of rapid
  adiabatic passage},\ }\href {https://doi.org/10.1103/PhysRevA.102.040201}
  {\bibfield  {journal} {\bibinfo  {journal} {Phys. Rev. A}\ }\textbf {\bibinfo
  {volume} {102}},\ \bibinfo {pages} {040201(R)} (\bibinfo {year}
  {2020})}\BibitemShut {NoStop}%
\bibitem [{\citenamefont {Sch{\"a}fer}\ \emph {et~al.}(2022)\citenamefont
  {Sch{\"a}fer}, \citenamefont {Budich},\ and\ \citenamefont {Luitz}}]{SBL22}%
  \BibitemOpen
  \bibfield  {author} {\bibinfo {author} {\bibfnamefont {R.}~\bibnamefont
  {Sch{\"a}fer}}, \bibinfo {author} {\bibfnamefont {J.~C.}\ \bibnamefont
  {Budich}},\ and\ \bibinfo {author} {\bibfnamefont {D.~J.}\ \bibnamefont
  {Luitz}},\ }\bibfield  {title} {\bibinfo {title} {Symmetry protected
  exceptional points of interacting fermions},\ }\href
  {https://arxiv.org/abs/2204.05340} {\bibfield  {journal} {\bibinfo  {journal}
  {arXiv:2204.05340}\ } (\bibinfo {year} {2022})}\BibitemShut {NoStop}%
\bibitem [{\citenamefont {Zheng}\ \emph {et~al.}(2015)\citenamefont {Zheng},
  \citenamefont {Hsu}, \citenamefont {Igarashi}, \citenamefont {Lu},
  \citenamefont {Kaminer}, \citenamefont {Pick}, \citenamefont {Chua},
  \citenamefont {Joannopoulos},\ and\ \citenamefont
  {Solja\u{c}i{\'c}}}]{ZHI15}%
  \BibitemOpen
  \bibfield  {author} {\bibinfo {author} {\bibfnamefont {B.}~\bibnamefont
  {Zheng}}, \bibinfo {author} {\bibfnamefont {C.~W.}\ \bibnamefont {Hsu}},
  \bibinfo {author} {\bibfnamefont {Y.}~\bibnamefont {Igarashi}}, \bibinfo
  {author} {\bibfnamefont {L.}~\bibnamefont {Lu}}, \bibinfo {author}
  {\bibfnamefont {I.}~\bibnamefont {Kaminer}}, \bibinfo {author} {\bibfnamefont
  {A.}~\bibnamefont {Pick}}, \bibinfo {author} {\bibfnamefont {S.-L.}\
  \bibnamefont {Chua}}, \bibinfo {author} {\bibfnamefont {J.~D.}\ \bibnamefont
  {Joannopoulos}},\ and\ \bibinfo {author} {\bibfnamefont {M.}~\bibnamefont
  {Solja\u{c}i{\'c}}},\ }\bibfield  {title} {\bibinfo {title} {Spawning rings
  of exceptional points out of {D}irac cones},\ }\href
  {https://doi.org/10.1038/nature14889} {\bibfield  {journal} {\bibinfo
  {journal} {Nature (London)}\ }\textbf {\bibinfo {volume} {525}},\ \bibinfo
  {pages} {354} (\bibinfo {year} {2015})}\BibitemShut {NoStop}%
\bibitem [{\citenamefont {Lin}\ \emph {et~al.}(2016)\citenamefont {Lin},
  \citenamefont {Pick}, \citenamefont {Loncar},\ and\ \citenamefont
  {Rodriguez}}]{LPL16}%
  \BibitemOpen
  \bibfield  {author} {\bibinfo {author} {\bibfnamefont {Z.}~\bibnamefont
  {Lin}}, \bibinfo {author} {\bibfnamefont {A.}~\bibnamefont {Pick}}, \bibinfo
  {author} {\bibfnamefont {M.}~\bibnamefont {Loncar}},\ and\ \bibinfo {author}
  {\bibfnamefont {A.~W.}\ \bibnamefont {Rodriguez}},\ }\bibfield  {title}
  {\bibinfo {title} {Enhanced spontaneous emission at third-order {D}irac
  exceptional points in inverse-designed photonic crystals},\ }\href
  {https://doi.org/10.1103/PhysRevLett.117.107402} {\bibfield  {journal}
  {\bibinfo  {journal} {Phys. Rev. Lett.}\ }\textbf {\bibinfo {volume} {117}},\
  \bibinfo {pages} {107402} (\bibinfo {year} {2016})}\BibitemShut {NoStop}%
\bibitem [{\citenamefont {Zhou}\ \emph {et~al.}(2018)\citenamefont {Zhou},
  \citenamefont {Peng}, \citenamefont {Yoon}, \citenamefont {Hsu},
  \citenamefont {Nelson}, \citenamefont {Fu}, \citenamefont {Joannopoulos},
  \citenamefont {Solja\u{c}i{\'c}},\ and\ \citenamefont {Zhen}}]{ZPY18}%
  \BibitemOpen
  \bibfield  {author} {\bibinfo {author} {\bibfnamefont {H.}~\bibnamefont
  {Zhou}}, \bibinfo {author} {\bibfnamefont {C.}~\bibnamefont {Peng}}, \bibinfo
  {author} {\bibfnamefont {Y.}~\bibnamefont {Yoon}}, \bibinfo {author}
  {\bibfnamefont {C.~W.}\ \bibnamefont {Hsu}}, \bibinfo {author} {\bibfnamefont
  {K.~A.}\ \bibnamefont {Nelson}}, \bibinfo {author} {\bibfnamefont
  {L.}~\bibnamefont {Fu}}, \bibinfo {author} {\bibfnamefont {J.~D.}\
  \bibnamefont {Joannopoulos}}, \bibinfo {author} {\bibfnamefont
  {M.}~\bibnamefont {Solja\u{c}i{\'c}}},\ and\ \bibinfo {author} {\bibfnamefont
  {B.}~\bibnamefont {Zhen}},\ }\bibfield  {title} {\bibinfo {title}
  {Observation of bulk {F}ermi arc and polarization half charge from paired
  exceptional points},\ }\href {https://doi.org/10.1126/science.aap9859}
  {\bibfield  {journal} {\bibinfo  {journal} {Nature (London)}\ }\textbf
  {\bibinfo {volume} {359}},\ \bibinfo {pages} {1009} (\bibinfo {year}
  {2018})}\BibitemShut {NoStop}%
\bibitem [{\citenamefont {Cerjan}\ \emph {et~al.}(2019)\citenamefont {Cerjan},
  \citenamefont {Huang}, \citenamefont {Wang}, \citenamefont {Chen},
  \citenamefont {Chong},\ and\ \citenamefont {Rechtsman}}]{CSHW19}%
  \BibitemOpen
  \bibfield  {author} {\bibinfo {author} {\bibfnamefont {A.}~\bibnamefont
  {Cerjan}}, \bibinfo {author} {\bibfnamefont {S.}~\bibnamefont {Huang}},
  \bibinfo {author} {\bibfnamefont {M.}~\bibnamefont {Wang}}, \bibinfo {author}
  {\bibfnamefont {K.~P.}\ \bibnamefont {Chen}}, \bibinfo {author}
  {\bibfnamefont {Y.}~\bibnamefont {Chong}},\ and\ \bibinfo {author}
  {\bibfnamefont {M.~C.}\ \bibnamefont {Rechtsman}},\ }\bibfield  {title}
  {\bibinfo {title} {Experimental realization of a {W}eyl exceptional ring},\
  }\href {https://doi.org/10.1038/s41566-019-0453-z} {\bibfield  {journal}
  {\bibinfo  {journal} {Nat. Photonics}\ }\textbf {\bibinfo {volume} {13}},\
  \bibinfo {pages} {623} (\bibinfo {year} {2019})}\BibitemShut {NoStop}%
\bibitem [{\citenamefont {Zyuzin}\ and\ \citenamefont {Simon}(2019)}]{ZS19}%
  \BibitemOpen
  \bibfield  {author} {\bibinfo {author} {\bibfnamefont {A.~A.}\ \bibnamefont
  {Zyuzin}}\ and\ \bibinfo {author} {\bibfnamefont {P.}~\bibnamefont {Simon}},\
  }\bibfield  {title} {\bibinfo {title} {Disorder-induced exceptional points
  and nodal lines in {D}irac superconductors},\ }\href
  {https://doi.org/10.1103/PhysRevB.99.165145} {\bibfield  {journal} {\bibinfo
  {journal} {Phys. Rev. B}\ }\textbf {\bibinfo {volume} {99}},\ \bibinfo
  {pages} {165145} (\bibinfo {year} {2019})}\BibitemShut {NoStop}%
\bibitem [{\citenamefont {Rausch}\ \emph {et~al.}(2021)\citenamefont {Rausch},
  \citenamefont {Peter},\ and\ \citenamefont {Yoshida}}]{RPY21}%
  \BibitemOpen
  \bibfield  {author} {\bibinfo {author} {\bibfnamefont {R.}~\bibnamefont
  {Rausch}}, \bibinfo {author} {\bibfnamefont {R.}~\bibnamefont {Peter}},\ and\
  \bibinfo {author} {\bibfnamefont {T.}~\bibnamefont {Yoshida}},\ }\bibfield
  {title} {\bibinfo {title} {Exceptional points in the one-dimensional
  {H}ubbard model},\ }\href {https://doi.org/10.1088/1367-2630/abd35e}
  {\bibfield  {journal} {\bibinfo  {journal} {New J. Phys.}\ }\textbf {\bibinfo
  {volume} {23}},\ \bibinfo {pages} {013011} (\bibinfo {year}
  {2021})}\BibitemShut {NoStop}%
\bibitem [{\citenamefont {Horn}\ and\ \citenamefont {Johnson}(2013)}]{HJ13}%
  \BibitemOpen
  \bibfield  {author} {\bibinfo {author} {\bibfnamefont {R.~A.}\ \bibnamefont
  {Horn}}\ and\ \bibinfo {author} {\bibfnamefont {C.~R.}\ \bibnamefont
  {Johnson}},\ }\href
  {https://www.cambridge.org/de/academic/subjects/mathematics/algebra/matrix-analysis-2nd-edition?format=PB&isbn=9780521548236}
  {\emph {\bibinfo {title} {Matrix Analysis}}}\ (\bibinfo  {publisher}
  {Cambridge University Press},\ \bibinfo {address} {Cambridge},\ \bibinfo
  {year} {2013})\BibitemShut {NoStop}%
\bibitem [{\citenamefont {Gentle}(2017)}]{Gentle17}%
  \BibitemOpen
  \bibfield  {author} {\bibinfo {author} {\bibfnamefont {J.~E.}\ \bibnamefont
  {Gentle}},\ }\href {https://link.springer.com/book/10.1007/978-0-387-70873-7}
  {\emph {\bibinfo {title} {Matrix Algebra: Theory, Computations, and
  Applications in Statistics}}}\ (\bibinfo  {publisher} {Springer},\ \bibinfo
  {address} {Switzerland},\ \bibinfo {year} {2017})\BibitemShut {NoStop}%
\bibitem [{\citenamefont {Higham}(1989)}]{Higham89}%
  \BibitemOpen
  \bibfield  {author} {\bibinfo {author} {\bibfnamefont {N.~J.}\ \bibnamefont
  {Higham}},\ }\bibfield  {title} {\bibinfo {title} {Matrix nearness problems
  and applications},\ }in\ \href
  {https://www.cambridge.org/core/journals/mathematical-gazette/article/abs/applications-of-matrix-theory-edited-by-m-j-c-gover-and-s-barnett-pp-324-40-1989-isbn-0198536259-oxford-university-press/28B0DC003EE9BFF5F1E3289006821F60}
  {\emph {\bibinfo {booktitle} {Applications of Matrix Theory}}},\ \bibinfo
  {editor} {edited by\ \bibinfo {editor} {\bibfnamefont {M.~J.~C.}\
  \bibnamefont {Gover}}\ and\ \bibinfo {editor} {\bibfnamefont
  {S.}~\bibnamefont {Barnett}}}\ (\bibinfo  {publisher} {Oxford University
  Press},\ \bibinfo {address} {Berlin},\ \bibinfo {year} {1989})\BibitemShut
  {NoStop}%
\bibitem [{\citenamefont {Demmel}(1987)}]{Demmel87}%
  \BibitemOpen
  \bibfield  {author} {\bibinfo {author} {\bibfnamefont {J.~W.}\ \bibnamefont
  {Demmel}},\ }\bibfield  {title} {\bibinfo {title} {On condition numbers and
  the distance to the nearest ill-posed problem},\ }\href
  {https://doi.org/10.1007/BF01400115} {\bibfield  {journal} {\bibinfo
  {journal} {Numer. Math.}\ }\textbf {\bibinfo {volume} {51}},\ \bibinfo
  {pages} {251} (\bibinfo {year} {1987})}\BibitemShut {NoStop}%
\bibitem [{\citenamefont {Chalker}\ and\ \citenamefont {Mehlig}(1998)}]{CM98}%
  \BibitemOpen
  \bibfield  {author} {\bibinfo {author} {\bibfnamefont {J.~T.}\ \bibnamefont
  {Chalker}}\ and\ \bibinfo {author} {\bibfnamefont {B.}~\bibnamefont
  {Mehlig}},\ }\bibfield  {title} {\bibinfo {title} {Eigenvector statistics in
  non-{H}ermitian random matrix ensembles},\ }\href
  {https://doi.org/10.1103/PhysRevLett.81.3367} {\bibfield  {journal} {\bibinfo
   {journal} {Phys. Rev. Lett.}\ }\textbf {\bibinfo {volume} {81}},\ \bibinfo
  {pages} {3367} (\bibinfo {year} {1998})}\BibitemShut {NoStop}%
\bibitem [{\citenamefont {Fyodorov}\ and\ \citenamefont {Mehlig}(2002)}]{FM02}%
  \BibitemOpen
  \bibfield  {author} {\bibinfo {author} {\bibfnamefont {Y.~V.}\ \bibnamefont
  {Fyodorov}}\ and\ \bibinfo {author} {\bibfnamefont {B.}~\bibnamefont
  {Mehlig}},\ }\bibfield  {title} {\bibinfo {title} {Statistics of resonances
  and nonorthogonal eigenfunctions in a model for single-channel chaotic
  scattering},\ }\href {https://doi.org/10.1103/PhysRevE.66.045202} {\bibfield
  {journal} {\bibinfo  {journal} {Phys. Rev. E}\ }\textbf {\bibinfo {volume}
  {66}},\ \bibinfo {pages} {045202(R)} (\bibinfo {year} {2002})}\BibitemShut
  {NoStop}%
\bibitem [{\citenamefont {Wiersig}(2019)}]{Wiersig19}%
  \BibitemOpen
  \bibfield  {author} {\bibinfo {author} {\bibfnamefont {J.}~\bibnamefont
  {Wiersig}},\ }\bibfield  {title} {\bibinfo {title} {Nonorthogonality
  constraints in open quantum and wave systems},\ }\href
  {https://doi.org/10.1103/PhysRevResearch.1.033182} {\bibfield  {journal}
  {\bibinfo  {journal} {Phys. Rev. Res.}\ }\textbf {\bibinfo {volume} {1}},\
  \bibinfo {pages} {033182} (\bibinfo {year} {2019})}\BibitemShut {NoStop}%
\bibitem [{\citenamefont {Schomerus}(2009)}]{Schomerus09}%
  \BibitemOpen
  \bibfield  {author} {\bibinfo {author} {\bibfnamefont {H.}~\bibnamefont
  {Schomerus}},\ }\bibfield  {title} {\bibinfo {title} {Excess quantum noise
  due to mode nonorthogonality in dielectric microresonators},\ }\href
  {https://doi.org/10.1103/PhysRevA.79.061801} {\bibfield  {journal} {\bibinfo
  {journal} {Phys. Rev. A}\ }\textbf {\bibinfo {volume} {79}},\ \bibinfo
  {pages} {061801(R)} (\bibinfo {year} {2009})}\BibitemShut {NoStop}%
\bibitem [{\citenamefont {Schomerus}(2017)}]{HS17}%
  \BibitemOpen
  \bibfield  {author} {\bibinfo {author} {\bibfnamefont {H.}~\bibnamefont
  {Schomerus}},\ }\bibfield  {title} {\bibinfo {title} {Random matrix
  approaches to open quantum systems},\ }in\ \href
  {https://doi.org/10.1093/oso/9780198797319.003.0010} {\emph {\bibinfo
  {booktitle} {Stochastic Processes and Random Matrices}}},\ \bibinfo {series}
  {Les Houches Summer School Lectures}, Vol.\ \bibinfo {volume} {104},\
  \bibinfo {editor} {edited by\ \bibinfo {editor} {\bibfnamefont
  {G.}~\bibnamefont {Schehr}} \emph {et~al.}}\ (\bibinfo  {publisher} {North
  Holland},\ \bibinfo {address} {Amsterdam},\ \bibinfo {year}
  {2017})\BibitemShut {NoStop}%
\bibitem [{\citenamefont {Trefethen}\ and\ \citenamefont
  {Embree}(2005)}]{TE05}%
  \BibitemOpen
  \bibfield  {author} {\bibinfo {author} {\bibfnamefont {L.~N.}\ \bibnamefont
  {Trefethen}}\ and\ \bibinfo {author} {\bibfnamefont {M.}~\bibnamefont
  {Embree}},\ }\href {https://press.princeton.edu/titles/8113.html} {\emph
  {\bibinfo {title} {Spectra and Pseudospectra}}}\ (\bibinfo  {publisher}
  {Princeton University Press},\ \bibinfo {address} {Princeton, NJ},\ \bibinfo
  {year} {2005})\BibitemShut {NoStop}%
\bibitem [{\citenamefont {Henrici}(1962)}]{Henrici62}%
  \BibitemOpen
  \bibfield  {author} {\bibinfo {author} {\bibfnamefont {P.}~\bibnamefont
  {Henrici}},\ }\bibfield  {title} {\bibinfo {title} {Bounds for iterates,
  inverses, spectral variation and fields of values of non-normal matrices},\
  }\href {https://eudml.org/doc/131513} {\bibfield  {journal} {\bibinfo
  {journal} {Numerische Mathematik}\ }\textbf {\bibinfo {volume} {4}},\
  \bibinfo {pages} {24} (\bibinfo {year} {1962})}\BibitemShut {NoStop}%
\bibitem [{\citenamefont {Gil}(2003)}]{Gil03}%
  \BibitemOpen
  \bibfield  {author} {\bibinfo {author} {\bibfnamefont {M.~I.}\ \bibnamefont
  {Gil}},\ }\href {https://link.springer.com/book/10.1007/b93845} {\emph
  {\bibinfo {title} {Operator Functions and Localization of Spectra}}},\
  \bibinfo {series} {Lecture Notes in Mathematics}, Vol.\ \bibinfo {volume}
  {1830}\ (\bibinfo  {publisher} {Springer},\ \bibinfo {address} {Berlin},\
  \bibinfo {year} {2003})\BibitemShut {NoStop}%
\bibitem [{\citenamefont {Wiersig}(2016)}]{Wiersig16}%
  \BibitemOpen
  \bibfield  {author} {\bibinfo {author} {\bibfnamefont {J.}~\bibnamefont
  {Wiersig}},\ }\bibfield  {title} {\bibinfo {title} {Sensors operating at
  exceptional points: General theory},\ }\href
  {https://doi.org/10.1103/PhysRevA.93.033809} {\bibfield  {journal} {\bibinfo
  {journal} {Phys. Rev. A}\ }\textbf {\bibinfo {volume} {93}},\ \bibinfo
  {pages} {033809} (\bibinfo {year} {2016})}\BibitemShut {NoStop}%
\bibitem [{\citenamefont {Sunada}(2018)}]{Sunada18}%
  \BibitemOpen
  \bibfield  {author} {\bibinfo {author} {\bibfnamefont {S.}~\bibnamefont
  {Sunada}},\ }\bibfield  {title} {\bibinfo {title} {Enhanced response of
  non-{H}ermitian photonic systems near exceptional points},\ }\href
  {https://doi.org/10.1103/PhysRevA.97.043804} {\bibfield  {journal} {\bibinfo
  {journal} {Phys. Rev. A}\ }\textbf {\bibinfo {volume} {97}},\ \bibinfo
  {pages} {043804} (\bibinfo {year} {2018})}\BibitemShut {NoStop}%
\bibitem [{\citenamefont {Zhong}\ \emph
  {et~al.}(2020{\natexlab{a}})\citenamefont {Zhong}, \citenamefont {Kou},
  \citenamefont {{\"O}zdemir},\ and\ \citenamefont {El-Ganainy}}]{ZKO20}%
  \BibitemOpen
  \bibfield  {author} {\bibinfo {author} {\bibfnamefont {Q.}~\bibnamefont
  {Zhong}}, \bibinfo {author} {\bibfnamefont {J.}~\bibnamefont {Kou}}, \bibinfo
  {author} {\bibfnamefont {{\c{S}}.~K.}\ \bibnamefont {{\"O}zdemir}},\ and\
  \bibinfo {author} {\bibfnamefont {R.}~\bibnamefont {El-Ganainy}},\ }\bibfield
   {title} {\bibinfo {title} {Hierarchical construction of higher-order
  exceptional points},\ }\href {https://doi.org/10.1103/PhysRevLett.125.203602}
  {\bibfield  {journal} {\bibinfo  {journal} {Phys. Rev. Lett.}\ }\textbf
  {\bibinfo {volume} {125}},\ \bibinfo {pages} {203602} (\bibinfo {year}
  {2020}{\natexlab{a}})}\BibitemShut {NoStop}%
\bibitem [{\citenamefont {Zhong}\ \emph
  {et~al.}(2020{\natexlab{b}})\citenamefont {Zhong}, \citenamefont
  {{\"O}zdemir}, \citenamefont {Eisfeld}, \citenamefont {Metelmann},\ and\
  \citenamefont {El-Ganainy}}]{ZOE20}%
  \BibitemOpen
  \bibfield  {author} {\bibinfo {author} {\bibfnamefont {Q.}~\bibnamefont
  {Zhong}}, \bibinfo {author} {\bibfnamefont {{\c{S}}.~K.}\ \bibnamefont
  {{\"O}zdemir}}, \bibinfo {author} {\bibfnamefont {A.}~\bibnamefont
  {Eisfeld}}, \bibinfo {author} {\bibfnamefont {A.}~\bibnamefont {Metelmann}},\
  and\ \bibinfo {author} {\bibfnamefont {R.}~\bibnamefont {El-Ganainy}},\
  }\bibfield  {title} {\bibinfo {title} {Exceptional points-based optical
  amplifiers},\ }\href {https://doi.org/10.1103/PhysRevApplied.13.014070}
  {\bibfield  {journal} {\bibinfo  {journal} {Phys. Rev. Appl.}\ }\textbf
  {\bibinfo {volume} {13}},\ \bibinfo {pages} {014070} (\bibinfo {year}
  {2020}{\natexlab{b}})}\BibitemShut {NoStop}%
\bibitem [{\citenamefont {N{\"o}ckel}\ and\ \citenamefont
  {Stone}(1997)}]{ND97}%
  \BibitemOpen
  \bibfield  {author} {\bibinfo {author} {\bibfnamefont {J.~U.}\ \bibnamefont
  {N{\"o}ckel}}\ and\ \bibinfo {author} {\bibfnamefont {A.~D.}\ \bibnamefont
  {Stone}},\ }\bibfield  {title} {\bibinfo {title} {Ray and wave chaos in
  asymmetric resonant optical cavities},\ }\href
  {https://doi.org/10.1038/385045a0} {\bibfield  {journal} {\bibinfo  {journal}
  {Nature (London)}\ }\textbf {\bibinfo {volume} {385}},\ \bibinfo {pages} {45}
  (\bibinfo {year} {1997})}\BibitemShut {NoStop}%
\bibitem [{\citenamefont {Gmachl}\ \emph {et~al.}(1998)\citenamefont {Gmachl},
  \citenamefont {Capasso}, \citenamefont {Narimanov}, \citenamefont
  {N{\"o}ckel}, \citenamefont {Stone}, \citenamefont {Faist}, \citenamefont
  {Sivco},\ and\ \citenamefont {Cho}}]{GCNNSFSC98}%
  \BibitemOpen
  \bibfield  {author} {\bibinfo {author} {\bibfnamefont {C.}~\bibnamefont
  {Gmachl}}, \bibinfo {author} {\bibfnamefont {F.}~\bibnamefont {Capasso}},
  \bibinfo {author} {\bibfnamefont {E.~E.}\ \bibnamefont {Narimanov}}, \bibinfo
  {author} {\bibfnamefont {J.~U.}\ \bibnamefont {N{\"o}ckel}}, \bibinfo
  {author} {\bibfnamefont {A.~D.}\ \bibnamefont {Stone}}, \bibinfo {author}
  {\bibfnamefont {J.}~\bibnamefont {Faist}}, \bibinfo {author} {\bibfnamefont
  {D.~L.}\ \bibnamefont {Sivco}},\ and\ \bibinfo {author} {\bibfnamefont
  {A.~Y.}\ \bibnamefont {Cho}},\ }\bibfield  {title} {\bibinfo {title}
  {High-{P}ower {D}irectional {E}mission from {M}icrolasers with {C}haotic
  {R}esonators},\ }\href {https://doi.org/10.1126/science.280.5369.1556}
  {\bibfield  {journal} {\bibinfo  {journal} {Science}\ }\textbf {\bibinfo
  {volume} {280}},\ \bibinfo {pages} {1556} (\bibinfo {year}
  {1998})}\BibitemShut {NoStop}%
\bibitem [{\citenamefont {Cao}\ and\ \citenamefont {Wiersig}(2015)}]{CW15}%
  \BibitemOpen
  \bibfield  {author} {\bibinfo {author} {\bibfnamefont {H.}~\bibnamefont
  {Cao}}\ and\ \bibinfo {author} {\bibfnamefont {J.}~\bibnamefont {Wiersig}},\
  }\bibfield  {title} {\bibinfo {title} {Dielectric microcavities: Model
  systems for wave chaos and non-{H}ermitian physics},\ }\href
  {https://doi.org/10.1103/RevModPhys.87.61} {\bibfield  {journal} {\bibinfo
  {journal} {Rev. Mod. Phys.}\ }\textbf {\bibinfo {volume} {87}},\ \bibinfo
  {pages} {61} (\bibinfo {year} {2015})}\BibitemShut {NoStop}%
\bibitem [{\citenamefont {N{\"o}ckel}\ \emph {et~al.}(1994)\citenamefont
  {N{\"o}ckel}, \citenamefont {Stone},\ and\ \citenamefont
  {Chang}}]{Noeckel94}%
  \BibitemOpen
  \bibfield  {author} {\bibinfo {author} {\bibfnamefont {J.~U.}\ \bibnamefont
  {N{\"o}ckel}}, \bibinfo {author} {\bibfnamefont {A.~D.}\ \bibnamefont
  {Stone}},\ and\ \bibinfo {author} {\bibfnamefont {R.~K.}\ \bibnamefont
  {Chang}},\ }\bibfield  {title} {\bibinfo {title} {{$Q$} spoiling and
  directionality in deformed ring cavities},\ }\href
  {https://doi.org/10.1364/OL.19.001693} {\bibfield  {journal} {\bibinfo
  {journal} {Opt. Lett.}\ }\textbf {\bibinfo {volume} {19}},\ \bibinfo {pages}
  {1693} (\bibinfo {year} {1994})}\BibitemShut {NoStop}%
\bibitem [{\citenamefont {Hatano}\ and\ \citenamefont {Nelson}(1996)}]{HN96}%
  \BibitemOpen
  \bibfield  {author} {\bibinfo {author} {\bibfnamefont {N.}~\bibnamefont
  {Hatano}}\ and\ \bibinfo {author} {\bibfnamefont {D.~R.}\ \bibnamefont
  {Nelson}},\ }\bibfield  {title} {\bibinfo {title} {Localization {T}ransitions
  in {N}on-{H}ermitian {Q}uantum {M}echanics},\ }\href
  {https://doi.org/10.1103/PhysRevLett.77.570} {\bibfield  {journal} {\bibinfo
  {journal} {Phys. Rev. Lett.}\ }\textbf {\bibinfo {volume} {77}},\ \bibinfo
  {pages} {570} (\bibinfo {year} {1996})}\BibitemShut {NoStop}%
\end{thebibliography}

\end{document}